%% file: main.tex
\documentclass[prb,reprint,nofootinbib]{revtex4-1}
\bibpunct{[}{]}{;}{n}{}{}

\pdfoutput=1
\usepackage[caption=false]{subfig}

\usepackage{graphicx}
\usepackage{comment}
\usepackage{amsmath}
\usepackage{amssymb}
\usepackage{amsfonts}
\usepackage{bbm}
\usepackage{braket}
\usepackage{verbatim}
\usepackage[hidelinks]{hyperref}
\usepackage{ulem}
\usepackage{color, colortbl}
\usepackage{multirow}
\definecolor{LightCyan}{rgb}{1,0.5,0.5}
\hypersetup{
  colorlinks   = true, 
  urlcolor     = blue, 
  linkcolor    = blue, 
  citecolor   = blue 
}

\usepackage{color}
\usepackage{tikz}

\newcommand{\maxcut}{\texttt{MAXCUT} }

\newcommand{\be}{\begin{equation}}
\newcommand{\ee}{\end{equation}}

\begin{document}

	\title{MAXCUT QAOA performance guarantees for $p>1$}
	
	\author{Jonathan Wurtz}
	\email[Corresponding author: ] {jonathan.wurtz@tufts.edu}
	\author{Peter Love}
	\affiliation{Department of Physics and Astronomy, Tufts University, Medford, Massachusetts 02155, USA}

	\date{2/2/2021}

	\begin{abstract}
   We obtain worst case performance guarantees for $p=2$ and $3$ QAOA for \texttt{MAXCUT} on $3$-regular graphs. Previous work by Farhi et al.~obtained a lower bound on the approximation ratio of $0.692$ for $p=1$. We find a lower bound of $0.7559$ for $p=2$, where worst case graphs are those with no cycles $\leq 5$. This bound holds for any 3 regular graph evaluated at particular fixed parameters. We conjecture a hierarchy for all $p$, where worst case graphs have with no cycles $\leq 2p+1$. Under this conjecture, the approximation ratio is at least $0.7924$ for all 3 regular graphs and $p=3$. In addition, using an indistinguishably argument we find an upper bound on the worst case approximation ratio for all $p$, which indicates classes of graphs for which there can be no quantum advantage for at least $p<6$.
	\end{abstract}

	\maketitle
    
    \section{Introduction}\label{sec:introduction}
    
    In the rapidly developing field of quantum technology, near term quantum devices \cite{Preskill2018} are the focus of much interest. Such noisy intermediate scale quantum (NISQ) devices lack error correction and have imperfect gate implementations and environmental isolation, which restrict them to implementing only low depth algorithms. Even with these constraints, can such a device display quantum advantage?
    
    One algorithm suitable for NISQ devices is the quantum approximate optimization algorithm (QAOA), a hybrid quantum classical combinatorial optimization algorithm \cite{farhi2014}. In QAOA, a classical computer optimizes $2p$ angles parameterizing an ansatz wavefunction by querying a near term quantum device. This wavefunction encodes an approximate solution to some combinatorial optimization problem. For $p\to\infty$, it is known that the ansatz wavefunction encodes the exact solution, which follows from the adiabatic theorem \cite{crooks2018}. For finite $p$, the picture is less clear. What $p$ is needed to outperform the best classical algorithm? Asking such questions leads to competition between quantum and classical algorithms \cite{hastings2019classical}. For example, a QAOA algorithm for E3LIN2 \cite{farhi201e3lin0} with quantum advantage was answered by an improved classical algorithm \cite{barak2015}, which prompted an improved QAOA version without advantage \cite{farhi201e3lin}.
    
    One can find worst case performance guarantees for particular classes of problem instances in QAOA. Approximate solutions to a problem achieve some fraction $C$ of the exact solution, called the approximation ratio. A worst case performance guarantee bounds this approximation ratio from below. If the minimum approximation ratio $C_\text{min}$ obtained from the quantum algorithm is larger than the value for the best classical algorithm, then the quantum algorithm has quantum advantage, as it will produce better approximate answers for all instances. It is important to ask what this worst case performance guarantee $C_\text{min}$ is for QAOA as a function of $p$.
    
    In this paper, we apply QAOA to the NP-hard graph partitioning problem of \texttt{MAXCUT} \cite{garey1976}, which partitions some graph into two sets by cutting a maximum number of edges. We will find that the worst case performance guarantee for 3-regular graphs and $p=2$ is $C_2\geq 0.7559$, confirming the observation of \cite{farhi2014} and improving on the original result for $p=1$ of $C_1\geq 0.692$, as expected. Under a conjecture that graphs with no ``visible" cycles are worst case, we find $C_3\geq 0.7924$ for $p=3$. Additionally, we use an argument where fixed $p$ QAOA cannot distinguish between large cycles of even and odd length to find an upper bound on expectation values, which upper bounds the minimum approximation ratio.
    
    The paper is structured as follows. Section \ref{sec:methods} reviews QAOA applied to the \texttt{MAXCUT} problem. Section \ref{sec:fixedP} and \ref{sec:AR_lower_bounds} details how expectation values and the approximation ratio can be computed efficiently for any bounded degree graph for fixed values of $p$.  Section \ref{sec:worst_case_p2} computes the worst case performance guarantee for the $p=1$ and $2$ cases. Sections \ref{sec:angle_choice} - \ref{sec:upper_bounds} discuss some of the implications of the worst case performance, and Section \ref{sec:conclusion} concludes with discussion and interesting future directions. 
    
    \section{The \texttt{MAXCUT} problem and QAOA}\label{sec:methods}
    
    The \maxcut problem is defined as follows. Given a graph $\mathcal G$ with vertices $V$ and edges $E$, the vertices are partitioned into two sets labeled by, say, $+$ or $-$. The goal is to find the partition of vertices such that a maximal number of edges have one vertex in each set. Restated, a solution to the \maxcut problem separates a graph $\mathcal G$ into two subgraphs by cutting the maximum number of edges.
    
    This problem is encoded in qubits as follows. For each vertex, assign a qubit. Given vertices $\langle i\rangle$ and edges $\langle ij\rangle$ for a graph $\mathcal G$, the maximum cut is given by the maximal eigenstate of the objective function
    
    \begin{equation}
        \hat C = \sum_{\langle ij\rangle}\frac{1}{2}\big(1-\hat \sigma_z^i\hat \sigma_z^j\big).
    \end{equation}

    Each term is a clause representing an edge of the graph $\mathcal G$, with an eigenvalue of $1$ if the edge is cut, and $0$ if the edge is not. Because $\hat C$ is made of a sum of commuting Pauli $\hat \sigma_z$ terms, any eigenstate is a product state, and the maximal state can be simply read out in the Z basis. The partitioning of vertices is obtained by assigning each vertex according to a Z measurement outcome $\pm1$.
    
    One method of computing an approximate maximal state of $\hat C$ is the quantum approximate optimization algorithm (QAOA) \cite{farhi2014,crosson2014,Wang2018,Willsch2020,larkin2020,Zhou2020,arute2020quantum}. QAOA optimizes a variational wave function by maximizing the expectation value of the objective function with respect to a set of parameters $\{\gamma\}$, $\{\beta\}$
    
    \begin{equation}\label{eq:min_HT_alphagamma}
    \begin{split}
        \quad F(\gamma,\beta) &= \langle \gamma,\beta | \hat C |\gamma,\beta\rangle,\\
         F_{\rm{max}}&= \underset{\gamma,\beta}{\texttt{MAX}}:\quad F(\gamma,\beta).\\
    \end{split}
    \end{equation}
    
    The state preparation and evaluation of the expectation value $\langle\hat C\rangle$ can be done on a small quantum device, while the optimization of variational parameters $\{\gamma\}$, $\{\beta\}$ can be performed on a classical computer. The QAOA ansatz wave function $|\gamma,\beta\rangle$ is defined as \cite{farhi2014}

    \begin{equation}
        |\gamma,\beta\rangle = e^{-i\beta_p\hat B}e^{-i\gamma_p \hat C}(\dots) e^{-i\beta_1\hat B}e^{-i\gamma_1\hat C}|+\rangle
    \end{equation}
    where $\hat B$, the ``mixing Hamiltonian", is defined as $\hat B=\sum_i\hat \sigma_x^i$ and $|+\rangle$ is the equally weighted superposition state or analogously the largest eigenstate of $\hat B$. Ellipses represent $p$ iterations of unitarily evolving the wavefunction alternatively with generators $\hat C$ and $\hat B$. In the limit $p\to\infty$ the optimal state $| \gamma,\beta\rangle$ approaches the exact maximal state \cite{farhi2014}.
    
    Given an approximate wavefunction with expectation value $\langle \hat C\rangle$, the state can be repeatedly measured $m$ times in the Z basis to find a bit string whose expectation value evaluates to at least $\langle \hat C\rangle(1-1/m)$ as an approximate $\texttt{MAXCUT}$ solution. This is due to the phenomena of concentration, wherein the variance of expectation values is much smaller then the expectation value itself \cite{larkin2020}.
    
    The approximation ratio for \maxcut is
    
    \begin{equation}\label{eq:approx_ratio_defn}
        C(\gamma,\beta)=\frac{F(\gamma,\beta)}{C_\text{max}}
    \end{equation}
    where $C_\text{max}$ is the maximum number of edges cut for a particular graph $\mathcal G$. A number in between 0 and 1 measures how close the variational state is to the exact maximal state. A larger number indicates better performance, as bitstrings from the measurement procedure will have a better \maxcut value. If $F_\text{max}=C_\text{max}$ then the variational state is the exact maximal state, and the approximation ratio is $1$. Note that in practice, $C_\text{max}$ may not be known, so bounding the approximation ratio from below requires $C_\text{max}$ to be bounded from above.
    
    \section{Fixed-$p$ algorithm}\label{sec:fixedP}
    
    It was found by Farhi et al.~in 2014 \cite{farhi2014} that for fixed graph degree $\nu$ and particular value of $p$, the numerical difficulty of simulating QAOA evolution grows at most doubly exponentially in $p$, and linearly with number of vertices $N$. In the interests of fixing notation and making the present paper self-contained let us begin by repeating the derivation of \cite{farhi2014} here. 
    
    The expectation value $F(\gamma,\beta)$ is
    \begin{align}
        F(\gamma,\beta) &= \sum_{\langle ij\rangle}f_{\langle ij\rangle}(\gamma,\beta)\nonumber\\
        \text{with\quad}f_{\langle ij\rangle}(\gamma,\beta)&=\frac{1}{2}\big\langle \gamma,\beta\big|1-\hat \sigma_z^i\hat \sigma_z^j\big|\gamma,\beta\big\rangle,
    \end{align}
    where the expectation value $F(\gamma,\beta)$ has been broken into a sum of terms $f_{\langle ij\rangle}(\gamma,\beta)$ corresponding to individual edges $\langle ij\rangle$. For a particular value of $p$, each value of $f_{\langle ij\rangle}(\gamma,\beta)$ may be computed as
    
    \begin{equation}
        \frac{1}{2} - \frac{1}{2}\langle +|(\dots)e^{i\gamma_p\hat C}e^{i\beta_p\hat B} \hat \sigma_z^i\hat \sigma_z^j e^{-i\beta_p\hat B}e^{-i\gamma_p\hat C}(\dots)|+\rangle
    \end{equation}
    where ellipses denote the action of the other $2p-2$ generators. 
    
    In the Heisenberg picture, this expectation value can be computed for any value of $N$. The first generator $\hat B$ rotates each objective function clause as
    
    \begin{align}
    \hat \sigma_z^i\hat \sigma_z^j\quad\to\quad(\cos(2\beta_p) \hat \sigma_z^i &+ \sin(2\beta_p)\hat \sigma_y^i)\times\nonumber\\
    &(\cos(2\beta_p) \hat \sigma_z^j + \sin(2\beta_p)\hat \sigma_y^j),
    \end{align}
    keeping the Heisenberg rotated operator local to the span of the two sites $i,j$. Terms $\hat \sigma_z^k\hat \sigma_z^l$ in the second generator $\hat C$ commute and cancel unless the edges $j,k$ overlap with $i$ or $j$. In that case, the $\hat \sigma_y$ are rotated into $\hat \sigma_x$ and $\hat \sigma_z$ by terms $\hat \sigma_z^k\hat\sigma_z^i$, growing to a span supporting 3 sites $i,j,k$ for terms such as $\hat\sigma_z^k\hat\sigma_y^i\hat \sigma_y^j$. Repeating this one layer deeper can rotate Pauli operators on $k$, and so forth. From this argument, it can be seen that after $p$ steps, the operator will have a support over a subgraph with vertices at most $p$ edges away from the initial vertices $i,j$.
    
    Given a graph $\mathcal G$ and edge $\langle ij\rangle$, in order to compute a value $f_{\langle ij\rangle}(\gamma,\beta)$ one may truncate the graph to an induced subgraph only including vertices which are at most $p$ edges away from either $i$ or $j$, and the presence of the other vertices does not contribute to the expectation value. We denote such a subgraph of edge $\langle ij\rangle$ within graph $\mathcal G$ to depth $p$ as $\mathcal G_{\langle ij\rangle}^p$. For a fixed $p$ and graph degree $\nu$, there are a finite number of unique subgraphs. For 3-regular graphs, where there are exactly $3$ edges incident on vertex, and for $p=1$ there are 3 subgraphs, with at most 6 vertices; for $p=2$, there are 123 subgraphs with at most 14 vertices; for $p=3$, there are 913,088 subgraphs with at most 30 vertices. See Appendix \ref{app:finding_graphs} for more details, and Table \ref{tab:graph_menengere} for enumerated subgraphs for $p=1$ and $2$. We will focus on 3-regular graphs, but these results generalize to other graphs with small bounded degree.

    As a technical note, because only vertices within $p$ steps of the edge $\langle ij\rangle$ need be considered to compute $f_{\langle ij\rangle}(\gamma,\beta)$, such expectation values may be efficiently calculated in the Schr\"odinger picture. Because a $\hat\sigma_z\hat\sigma_z$ operator only spreads to a span over the subgraph, one need only apply unitary operators over the subgraph. If the wavefunction is evolved in the Schr\"odinger picture under these unitaries, the state on all other sites remains an unentangled $|+\rangle$ product state, and thus one can consider the wavefunction only acting on the reduced Hilbert space of the $n$ vertices of the subgraph. This allows computation with the order $2^{n}$ values of the wavefunction on the subgraph, instead of the order $4^{n}$ values from a general operator acting on the subgraph in the Heisenberg picture, or order $2^N$ values of the wavefunction on the entire graph.

    \begin{figure}
        \centering
        \includegraphics[width=0.85\linewidth]{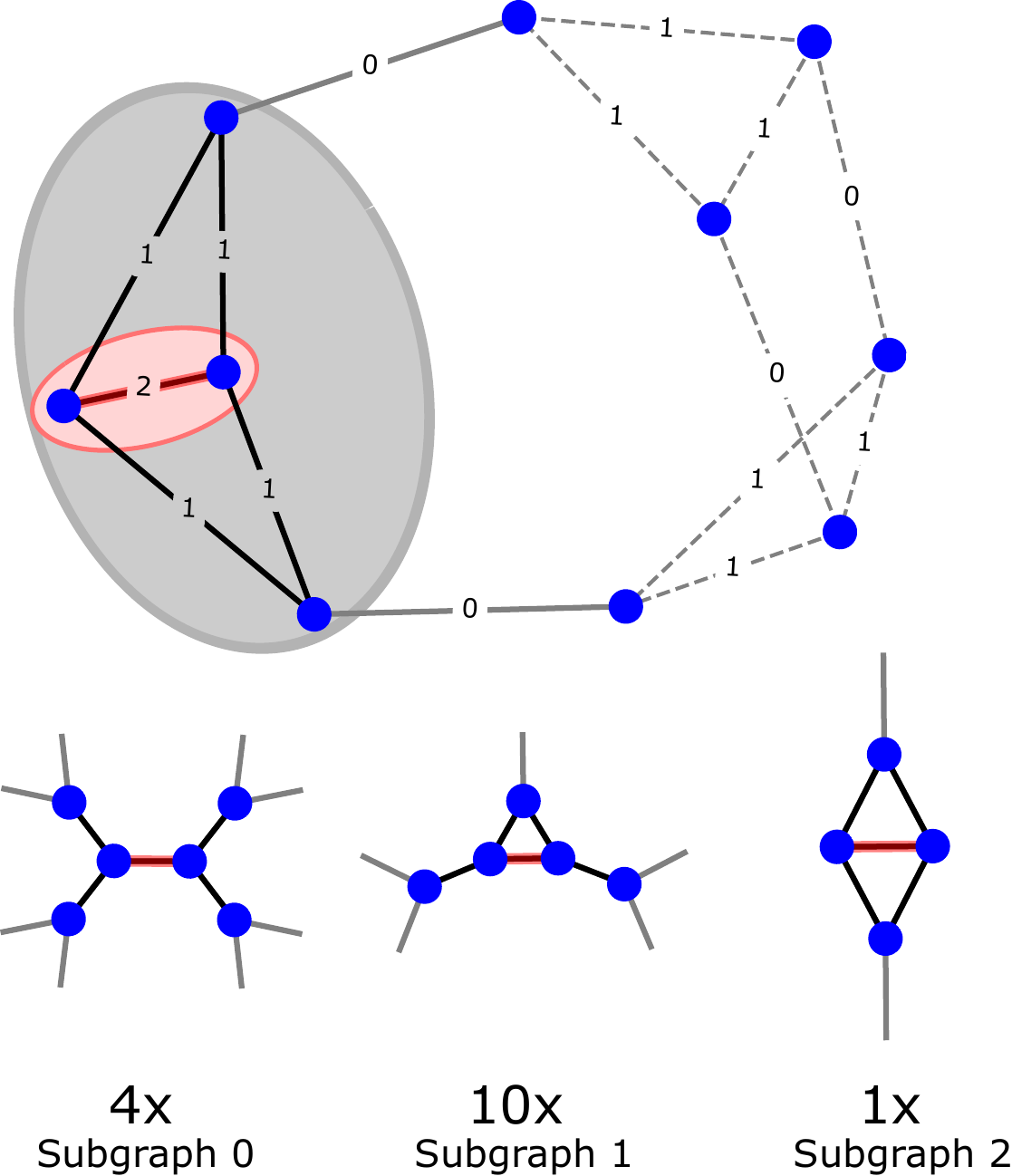}
        \caption{An example graph with subgraph assignments for $p=1$. Each edge (red) exists in a subgraph of edges and vertices a distance $p\leq 1$ away (grey circle). Edges within 1 step of the red edge (black) uniquely define the subgraph assignment of the edge. For this graph, there are 4 instances of subgraph 0 (``the tree"), 10 instances of subgraph 1 (``single triangle") and one instance of subgraph 2 (``two triangles"). The subgraph of each edge is labeled on the edge. A more detailed visual of this decomposition is shown in Fig.~\ref{fig:H_decomposition}.}
        \label{fig:example_tiling}
    \end{figure}

    The procedure for computing the expectation value $F(\gamma,\beta)$ for a particular graph $\mathcal G$ of bounded degree $\nu$ and fixed $p$ is as follows. For each edge $\langle ij\rangle$, identify the subgraph $\mathcal G_{\langle ij\rangle}^p$ of all edges and vertices within $p$ steps of $i$ and $j$ (See Fig.~\ref{fig:example_tiling}). This defines a collection of subgraphs $\{\mathcal G_{\langle ij\rangle}^p |\langle ij\rangle \in \mathcal G\}$, one for each edge, for which each $f_{\langle ij\rangle}(\gamma,\beta)$ can be computed in parallel. This collection of subgraphs can be further decomposed by counting the number $N_\lambda( \mathcal G)$ of each kind of subgraph $\lambda$, $\mathcal S_\lambda \in \{\mathcal S\}$ in the collection of all subgraphs of depth $p$, with each edge of the graph given a particular subgraph assignment $\langle ij\rangle\to \lambda$. The expectation value is then
    
    \begin{equation}\label{eq:approx_ratio_sum}
        F(\gamma,\beta) = \sum_{\text{subgraphs $\lambda$}}N_\lambda(\mathcal G)f_\lambda(\gamma,\beta)
    \end{equation}
    where $f_\lambda(\gamma,\beta)$ is the expectation value of the center edge of the $\lambda$th subgraph.
    
    
    \section{Lower bounds on the the approximation ratio}\label{sec:AR_lower_bounds}
    
    A performance guarantee for QAOA can be obtained by computing a lower bound of the approximation ratio $C_p(\mathcal G)=F_{M}/C_M$ of any graph, then finding the graph(s) with the lowest lower bound. Such a lower bound is given by the ratio of a lower bound on the maximum expectation value $F_{M}$, and an upper bound on the best \maxcut value $C_M$.
    
    \subsection{Lower bound on the maximum expectation value $F_M$}
    
    The value $F_{M}$ is bounded from below by:
    \begin{equation}\label{fmax}
        F_{M}=\underset{\gamma,\beta}{\texttt{MAX}}: F(\gamma,\beta) \quad\geq\quad \sum_\lambda N_\lambda(\mathcal G)f_\lambda
    \end{equation}
    where $f_\lambda\equiv f_\lambda(\gamma,\beta)$ is the expectation value of the center edge of the particular subgraph $\lambda$, chosen for a particular set of values $(\gamma,\beta)$. 
    
    The sum on the right hand side of eq.~(\ref{fmax}) is guaranteed to be less than or equal to the global maximum $F_\text{max}$, which simultaneously optimizes $(\gamma,\beta)$ for all clauses. We may choose the set of values for which to compute $f_\lambda$; we use the following set of angles
    \begin{align}
        p=1:\quad&\{\gamma_1,\beta_1\}=\{35.3^\circ,22.5^\circ\},\label{eq:angle_choices}\\
        p=2:\quad&\{\gamma_1,\beta_1,\gamma_2,\beta_2\}=\{28.0^\circ,31.8^\circ,51.4^\circ,16.8^\circ\}\nonumber.
    \end{align}
    These values are one of the optima for the tree subgraph, which does not have any cycles (see Fig. \ref{fig:example_tiling} bottom left). For more details on this choice of angles, see Section \ref{sec:angle_choice}.
    
    \begin{figure*}
        \centering
        \includegraphics{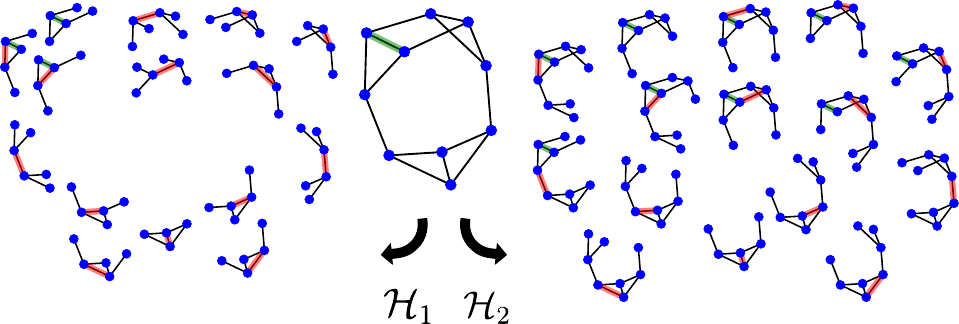}
        \caption{A lower bound on the approximation ratio can be found by a decomposition into subgraphs. Any graph $\mathcal G$ (center) can be decomposed into the subgraph graph $\mathcal H_p$. A lower bound on the expectation value is given by a sum over all subgraphs (Eq.~\eqref{eq:approx_ratio_sum}) where the center edge is in red. An upper bound on the \maxcut value is found by careful counting of uncut edges. A particular edge (green) appears five times in $\mathcal H_1$ (left) and nine times in $\mathcal H_2$ (right). A subgraph contributes to the sum~\eqref{eq:MAXCUT_env_condition} if the green edge participates in an odd length cycle of a subgraph. For this particular graph environment, 3 of 5 subgraphs contribute for $p=1$ and all 9 subgraphs contribute for $p=2$; Eq.~\eqref{eq:MAXCUT_env_condition} is 3/5 for $p=1$ and $452/495$ for $p=2$.}
        \label{fig:H_decomposition}
    \end{figure*}
    
    \subsection{Upper bound on the \maxcut value $C_M$}
    
    It is hard to find the exact \maxcut value $C_M$, which is after all one of the objectives of the QAOA algorithm. Fortunately, it suffices to find an upper bound of $C_M$ to yield a lower bound on the approximation ratio. Equivalently, we may find a \textit{lower} bound on the number of \textit{uncut} edges in a partition of a graph, $R_M(\mathcal G)$, to find the upper bound on the number of cut edges $C_M = N(\mathcal G) - R_M(\mathcal G)$. While these two views are equivalent, we find it more convenient to count uncut edges.

    The lower bound on the number of uncut edges can be found by considering only local structure of a graph. Locally, there is some amount of ``visible" frustration, which force a minimum number of edges to remain uncut. For example, a triangle of three vertices requires at least one edge uncut. Additional edges may remain uncut due to global structure which is not ``visible" locally.  We will use the local structure of a graph to get the tightest possible lower bound on the number of uncut edges. This will be done in three steps, each of which tighten the bound.
    
    As a first step, a trivial underestimate of the number of uncut edges is that no edges remain uncut in the graph, and so the \maxcut value is bounded from above by the number of edges in the graph. This bound is not tight, and does not take into account any of the structure of the graph.This trivial bound can be tightened by considering local structure. 
    
    As a next step, consider for a graph $\mathcal G$, a graph $\mathcal H_p$ which is a collection of disconnected subgraphs $\mathcal G_{\langle ij\rangle}^p$, one for each edge $\langle ij\rangle$ in $\mathcal G$. An example of this decomposition is shown in Fig.~\ref{fig:H_decomposition}. Heuristically, the graph $\mathcal H_p$ ``sees" the local structure of $\mathcal G$ out to a distance $p$. 
    
    Each edge $\langle ij\rangle$ in $\mathcal G$ appears in multiple subgraphs of $\mathcal G$ and therefore of $\mathcal H_p$. It appears in the subgraph $\mathcal G_{\langle ij\rangle}^p$ as the center edge, but it also appears in all subgraphs $\mathcal G_{\langle kl\rangle}^p$ whose center edge $\langle kl\rangle$ is $\leq p$ steps from edge $\langle ij\rangle$. Because distance is symmetric, the number of edges in subgraph $\mathcal G_{\langle ij\rangle}^p$ is equal to the number of subgraphs in which $\langle ij\rangle$ appears.
    
    The subgraphs in which edge $\langle ij\rangle$ appears are thus identified by all edges within a distance $2p$. The first $p$ steps identify the subgraph assignment of the center edge, while the second $p$ steps identify the subgraph assignment of the edges within $p$ steps of $\langle ij\rangle$.

    We call the surroundings $\mathcal G_{\langle ij\rangle}^{2p}$ which fix the subgraph assignment of adjacent edges to depth $p$ the {\em graph environment} of edge $\langle ij\rangle$. The set of all possible graph environments of depth $p$ is the set of all unique combinations of subgraph assignments on the edges of all subgraphs of depth $p$. It is also all possible combinations of subgraphs that an edge $\langle ij\rangle$ can appear in. This set of graph environments allows for a search through all possible local graph structures, without needing to be concerned with the global structure of the arbitrarily larger graph. 
    
\begin{figure}
        \centering
        \includegraphics[width=\linewidth]{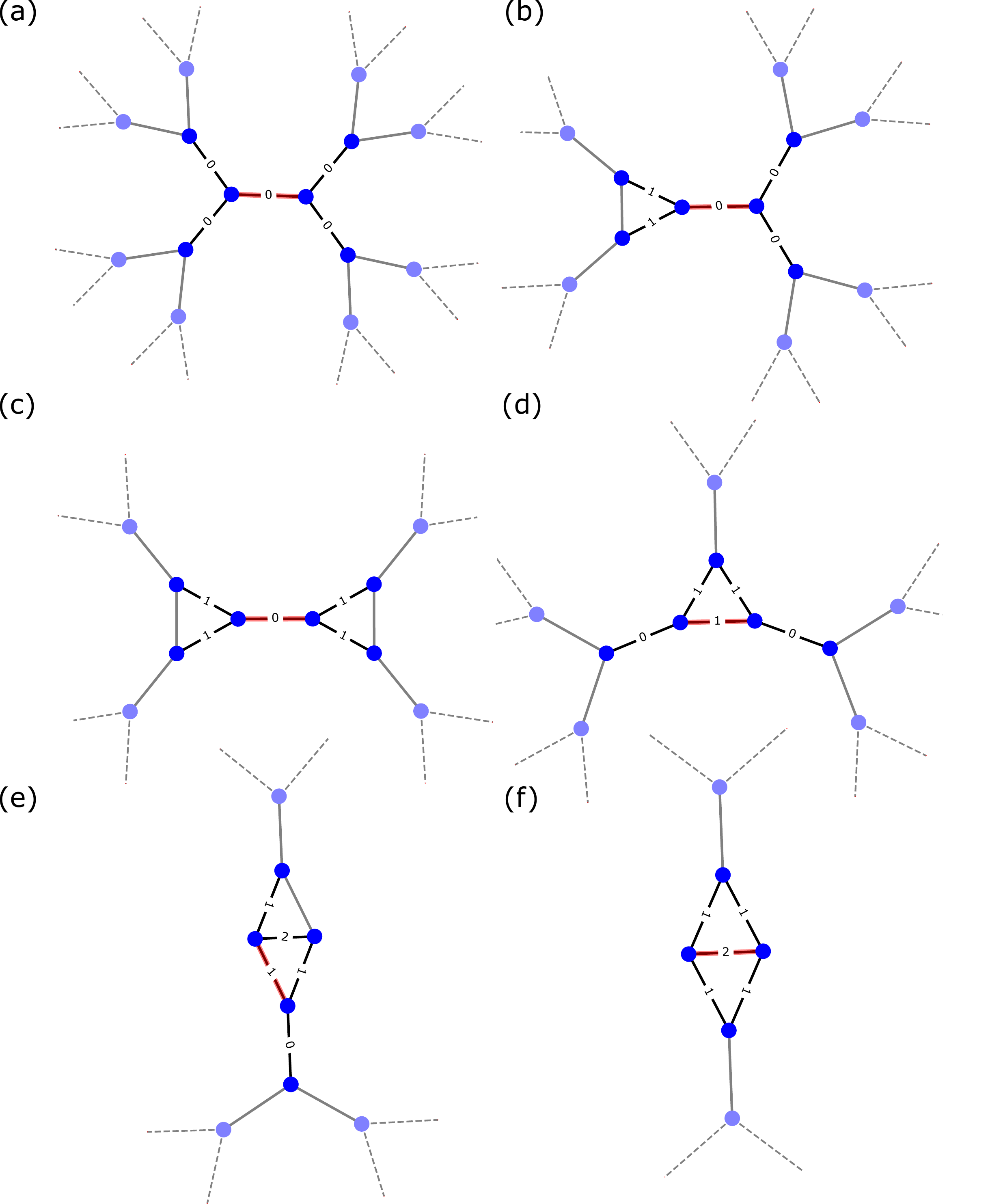}
        \caption{The six $p=1$ graph environments, which fix edges to particular subgraphs, as shown by edge labels. Black edges are of the center subgraph, while grey edges are choice of subgraph environment, and red is the special center edge. These graph environments mean that certain subgraphs cannot appear in isolation, and there are adjacency restrictions for certain subgraphs.}
        \label{fig:subgraph1_graph_envs}
    \end{figure}

    The six subgraph environments for $p=1$ are shown in Fig.~\ref{fig:subgraph1_graph_envs}. There are three combinations of subgraph assignments for subgraph 0, ``the tree" (Fig. \ref{fig:subgraph1_graph_envs}a,b,c), two combinations of subgraph assignments for subgraph 1, ``single triangle" (Fig.~\ref{fig:subgraph1_graph_envs}d,e), and one combination for subgraph 2, ``two triangles" (Fig.~\ref{fig:subgraph1_graph_envs}f).

\begin{figure}
        \centering
        \includegraphics{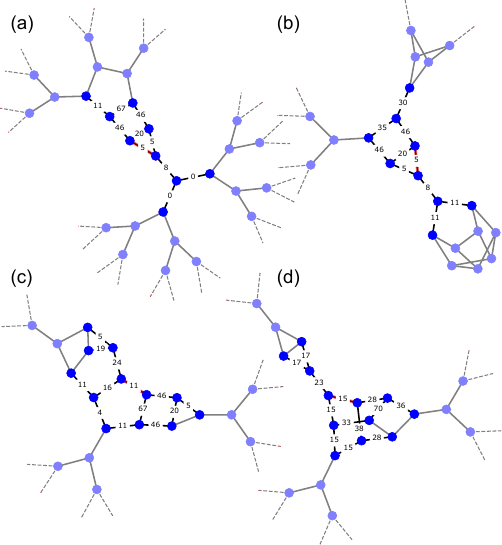}
        \caption{Some example graph environments for $p=2$ graphs. Subgraph edges fixed by the environment are shown by edge labels. (a,b) show two example environments for subgraph 5, while (c,d) show some example environments for subgraphs 11 and 15.}
        \label{fig:configs2}
    \end{figure}

    These graph environments restrict which subgraph assignments are allowed to be adjacent. For instance, consider subgraph 2, ``two triangles". There is only one graph environment in which it is the center edge, and the adjacent edges must be assigned to subgraph 1, ``one triangle". Thus, every instance of subgraph 2 in a graph must have at least 4 instances of subgraph 1. Equivalently, the set of graph environments are the set of all possible combinations of subgraphs which an edge can appear in. Some example graph environments for $p=2$ are shown in Fig.~\ref{fig:configs2}.

    The set of graph environments is a subset of all subgraphs of depth $2p$. It is a subset because subgraphs of depth $2p$ may have cycles of length more than $2p+1$, which the subgraph assignements of depth $p$ cannot distinguish. Because of this fact, the set of all graph environments is equivalent to the set of subgraphs of depth $2p$, subject to the constraint of no cycles greater than $2p+1$ in the minimum cycle basis~\cite{Paton1969}.

    Now, consider any partition $P$ of vertices of the  of graph $\mathcal G$, including the \texttt{MAXCUT} partition $M$. The partition leaves some edges $\langle ij\rangle$ uncut, denoted by the characteristic function $r_P$
    
    \begin{equation*}
    {r_P}_{\langle ij\rangle}=
        \begin{cases}
    1,& \text{if } \text{edge $\langle ij\rangle$ is uncut in the partition $P$ of $\mathcal G$}\\
    0,              & \text{otherwise.}
    \end{cases}
    \end{equation*}
    The number of uncut edges in $\mathcal G$ given partition $P$ is the sum over edges $R_P(\mathcal G) = \sum_{\langle ij\rangle} {r_P}_{\langle ij\rangle}$.
    
    Similarly, consider any partition ${\mathcal P}$ of vertices of the graph $\mathcal H_p$, including its \texttt{MAXCUT} partition ${\mathcal M}$. An edge of $\mathcal H_p$ is labeled by the index of its parent edge $\langle ij\rangle$ in $G$, as well as the center edge $\langle kl\rangle$ of the subgraph it participates in. The partition ${\mathcal P}$ leaves some edges $\langle ij\rangle$ of the subgraph $\mathcal G_{\langle kl\rangle}^p$ uncut. These cuts are labeled by the characteristic function $r_{\mathcal P}$
    
    \begin{equation*}
    {r_\mathcal P}_{\langle ij\rangle}^{\langle kl\rangle}=
        \begin{cases}
    1,& \text{if } \text{edge $\langle ij\rangle$ is uncut in  the partition}\\
    &\text{${\mathcal P}$ of subgraph $\mathcal G_{\langle kl\rangle}^p$}\\
    0,&              \text{if edge $\langle ij\rangle$ is cut in  the partition}\\
    &\text{${\mathcal P}$ of subgraph $\mathcal G_{\langle kl\rangle}^p$ or if $\langle ij\rangle\not \in \mathcal G_{\langle kl\rangle}^p$.}
    \end{cases}
    \end{equation*}
    The number of uncut edges in $\mathcal H_p$ given partition ${\mathcal P}$ is the sum over edges
    
    \begin{equation*}
    R_{\mathcal P}({\mathcal H}_p) = \sum_{\langle ij\rangle}\sum_{\langle kl\rangle} {r_{\mathcal P}}_{\langle ij\rangle}^{\langle kl\rangle}. 
    \end{equation*}

    The \maxcut $M$ of $\mathcal G$ induces a partition of $\mathcal H_p$, ${M}\mapsto {\mathcal I}$, in which each vertex of $\mathcal H_p$ is assigned to the same set as its parent vertex in $\mathcal G$. This implies that all copies of cut edges in $M$ of $\mathcal G$ are cut in $\mathcal H_p$, eg ${r_{\mathcal I}}_{\langle ij\rangle}^{\langle kl\rangle} = {r_\mathcal M}_{\langle ij\rangle}~\forall \langle kl\rangle \in {\mathcal G}$.
    
    The set of cuts of $\mathcal H_p$ is a superset of the cuts of $\mathcal G$ because the vertices of $\mathcal G$ appear multiple times in $\mathcal H_p$. The number of uncut edges in the \maxcut of $\mathcal H_p$ is therefore bounded above by the number of uncut edges in the partition ${\mathcal I}$ induced by the \maxcut of $\mathcal G$: $R_{\mathcal M}(\mathcal H_p)\leq R_{\mathcal I}(\mathcal H_p)$.
    
    Additionally, each edge $\langle ij\rangle$ of $\mathcal G$ appears $N(\mathcal G_{\langle ij\rangle}^p)$ times in $\mathcal H_p$. This number is bounded from above by the largest possible subgraph $N(\mathcal S_\lambda^p)$. For $p=1$, every subgraph has exactly $5$ edges. For $p=2$, the largest subgraphs have $13$ edges.
    
    Using the fact that the number of times each edge $\langle ij\rangle$ appears in $\mathcal H_p$ is bounded, and solutions on $\mathcal H_p$ can be induced from $\mathcal G$, we may bound the number of uncut edges in the \maxcut of $\mathcal G$ from below. For $p=1$ we may write the following inequalities
    \begin{align}
        &R_\mathcal{M}(\mathcal H_1)\leq R_\mathcal{I}(\mathcal H_1)\nonumber=\sum_{\langle ij\rangle,\langle kl\rangle}{r_\mathcal{I}}_{\langle ij\rangle}^{\langle kl\rangle}\\
        &=\sum_{\langle ij\rangle,\langle kl\rangle}{r_M}_{\langle ij\rangle}=5\sum_{\langle ij\rangle}{r_M}_{\langle ij\rangle}\nonumber
        \\& \Rightarrow\quad\frac{1}{5}R_\mathcal{M}(\mathcal H_1)\leq R_M(\mathcal G).\label{eq:p1_MAXCUT_bound1}
    \end{align}
    Similarly, for $p=2$,
    \begin{align}\label{eq:p2_MAXCUT_bound1}
        &R_\mathcal{M}(\mathcal H_2)\leq R_\mathcal{I}(\mathcal H_2)\nonumber=\sum_{\langle ij\rangle,\langle kl\rangle}{r_\mathcal{I}}_{\langle ij\rangle}^{\langle kl\rangle}\\
        &=\sum_{\langle ij\rangle,\langle kl\rangle}{r_M}_{\langle ij\rangle}\leq 13\sum_{\langle ij\rangle}{r_M}_{\langle ij\rangle}\nonumber
        \\& \Rightarrow\quad\frac{1}{13}R_\mathcal{M}(\mathcal H_2)\leq R_M(\mathcal G).
    \end{align}
    
    Because $\mathcal H_p$ is separated into disconnected subgraphs, it is simple to find the \maxcut partition of $\mathcal H_p$ for any size graph. This method of computing upper bounds on the \maxcut value reproduces that of \cite{farhi2014}. The bound on the \maxcut used in~ \cite{farhi2014} counted the number $T$ of isolated triangles (``single triangle" subgraphs) and $S$ of crossed squares (``two triangle" subgraphs). Each isolated triangle and crossed square will have one uncut edge, so a $3$-regular graph with $n$ vertices and $3n/2$ edges has at least $S+T$ uncut edges and thus at most $3n/2-S-T$ cut edges. Similarly, the number of $p=1$ subgraphs present in $\mathcal H_1$ are then functions of $S$ and $T$. Specifically, there are $S$ subgraphs of type 2 (``two triangle"), and $4S+3T$ subgraphs of type 1 (``single triangle"). Each of these subgraphs has one uncut edge, and so Eq.~\eqref{eq:p1_MAXCUT_bound1} upper bounds the maxcut of $\mathcal G$ to $3n/2 - (S + 4S + 3T)/5$. This bound is looser than that obtained from directly counting $S$ and $T$ in $\mathcal G$ because $H_1$ mistakes some of the crossed squares for isolated triangles. For $p=2$ the number of edges per subgraph in $H_2$ varies and use of Eq.~\eqref{eq:p2_MAXCUT_bound1} will give looser bounds. 
    
\begin{figure}[h!]
    \begin{center}
    \includegraphics[]{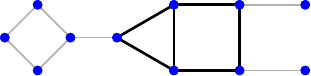}
    \end{center}
    \caption{An example of the function $\delta_{\langle ij\rangle}^{\langle kl\rangle}$ on an example graph. Here, $\delta=1$ for dark edges, which participate in odd cycles of length $3$ or $5$. Similarly, $\delta=0$ for light edges which do not participate in odd length cycles. This can occur even if the graph has odd length cycles, because parts of the graph are connected by only a single edge.\label{oddcycle}}
    \end{figure}

    The third and final step for finding a tighter bound for $p>1$ can be found by considering the local structure of subgraphs $G_{\langle kl\rangle}^p$. This requires the following fact: if an edge of some graph $\mathcal G$ participates in no odd length cycles, it must be cut in the \maxcut solution. This is a consequence of balance in signed graphs \cite{ZASLAVSKY2018}. This condition is labeled by the characteristic function:
    \begin{equation*}
    \delta_{\langle ij\rangle}^{\langle kl\rangle}=
        \begin{cases}
    1,& \text{if } \text{edge $\langle ij\rangle$ in subgraph $\mathcal G_{\langle kl\rangle}^p$ }\\&\text{participates in at least one odd length cycle}\\
    0,              & \text{otherwise.}
    \end{cases}
    \end{equation*}
    An example of this function on a graph is shown in Fig.~\ref{oddcycle}.   

     For large connected graphs, $\delta=1$ for almost every edge, suggesting this function is not very interesting for graphs with few ``loose edges", where a loose edge is defined as an edge connecting a vertex of degree one to the graph. However, the subgraphs $G_{\langle kl\rangle}^p$ have many loose edges (See Table~\ref{tab:graph_menengere}). Given some subgraph $G_{\langle kl\rangle}^p$  with \maxcut solution $\mathcal M$, this implies ${r_\mathcal{M}}_{\langle ij\rangle}^{\langle kl\rangle}={r_\mathcal{M}}_{\langle ij\rangle}^{\langle kl\rangle}{\delta}_{\langle ij\rangle}^{\langle kl\rangle}$.
    
    With this fact, let us follow the same procedure as above, inducing a solution $\mathcal I$ on subgraph $G_{\langle kl\rangle}^p$ from the \maxcut solution $M$ on graph $\mathcal G$, with ${r_\mathcal{I}}_{\langle ij\rangle}^{\langle kl\rangle}={r_M}_{\langle ij\rangle}$. It is simple to see that there exists a different partition $\mathcal{I}'$ for which
    
    \begin{align}
        {r_{\mathcal{I}'}}_{\langle ij\rangle}^{\langle kl\rangle} = {r_\mathcal{I}}_{\langle ij\rangle}^{\langle kl\rangle}\quad&\text{when}\quad\delta_{\langle ij\rangle}^{\langle kl\rangle}=1,\nonumber\\
        {r_{\mathcal{I}'}}_{\langle ij\rangle}^{\langle kl\rangle} = 0\qquad\quad &\text{when}\quad{\delta}_{\langle ij\rangle}^{\langle kl\rangle}=0.\nonumber
    \end{align}
    
    The partition $\mathcal{I}'$ exists based on the fact that $\delta=0$ only on ``loose edges" of a subgraph that are only connected by a single edge, and thus the loose edges can be solved independently from the rest of the subgraph.
    
    These facts lead to a chain of inequalities, where the sum is partitioned into edges which participate in odd length cycles, and those that do not
    
    \begin{multline}\label{eq:app:cut_edge_inequalities}
        R_\mathcal{I}(\mathcal G_{\langle kl\rangle}^p)=\sum_{\langle ij\rangle}\left({r_I}_{\langle ij\rangle}^{\langle kl\rangle}\delta_{\langle ij\rangle}^{\langle kl\rangle} + {r_I}_{\langle ij\rangle}^{\langle kl\rangle}(1-\delta_{\langle ij\rangle}^{\langle kl\rangle})\right)\\
        \geq R_{\mathcal{I}'}(\mathcal G_{\langle kl\rangle}^p)= \sum_{\langle ij\rangle}\left({r_{I'}}_{\langle ij\rangle}^{\langle kl\rangle}\delta_{\langle ij\rangle}^{\langle kl\rangle} + {r_{I'}}_{\langle ij\rangle}^{\langle kl\rangle}(1-\delta_{\langle ij\rangle}^{\langle kl\rangle})\right)\\
        =\sum_{\langle ij\rangle}{r_M}_{\langle ij\rangle}\delta_{\langle ij\rangle}^{\langle kl\rangle}\\
        \geq \sum_{\langle ij\rangle}{r_\mathcal{M}}_{\langle ij\rangle}^{\langle kl\rangle}=R_\mathcal{M}(\mathcal G_{\langle kl\rangle}^p).
    \end{multline}
    
    The first step is from the partition $\mathcal{M}'$ having a larger \maxcut than $\mathcal{M}$; the second step is from properties and definitions of the partition $\mathcal{M}'$ as induced from $\mathcal{M}$ as induced from $M$. The third step is from the \maxcut partition $\mathcal{M}$ being larger than the partition $\mathcal{M}'$.

    Next consider the following inequality:
    \begin{equation}\label{eq:uncut_lowerbound}
        \sum_{\langle kl\rangle}\frac{\sum_{\langle ij\rangle}{r_M}_{\langle ij\rangle}^{\langle kl\rangle}}{N(\mathcal G_{\langle kl\rangle}^p)}
        \leq 
        \sum_{\langle ij\rangle}{r_M}_{\langle ij\rangle}\left(\sum_{\langle kl\rangle}\frac{\delta_{\langle ij\rangle}^{\langle kl\rangle}}{N(\mathcal G_{\langle kl\rangle}^p)}\right),
    \end{equation}
    which follows from Eq.~\eqref{eq:app:cut_edge_inequalities}. It gives a lower bound on a weighted sum over edges present in the \maxcut of $G$. The maximum coefficient in the parenthesis is
     \begin{equation}\label{eq:MAXCUT_env_condition}
        \underset{\langle ij\rangle}{\texttt{MAX}}:\sum_{\langle kl\rangle}\frac{\delta_{\langle ij\rangle}^{\langle kl\rangle}}{N(\mathcal G_{\langle kl\rangle}^p)}= \frac{1}{\mu_p}
    \end{equation}
    for some yet undetermined factor $\mu_p$. This term gives the contribution from the worst-case graph environment and implies:
    \begin{equation}\label{eq:uncut_lowerbound_chain}
        \sum_{\langle kl\rangle}\frac{\sum_{\langle ij\rangle}{r_M}_{\langle ij\rangle}^{\langle kl\rangle}}{N(\mathcal G_{\langle kl\rangle}^p)}
        \leq \sum_{\langle ij\rangle}\sum_{\langle kl\rangle}\frac{{r_M}_{\langle ij\rangle}\delta_{\langle ij\rangle}^{\langle kl\rangle}}{N(\mathcal G_{\langle kl\rangle}^p)}
        \leq
        \frac{1}{\mu_p}\sum_{\langle ij\rangle} {r_M}_{\langle ij\rangle}.
    \end{equation}

    It thus suffices to search through every possible graph environment and thus every possible combination of subgraphs an edge can participate in to find the worst-case graph environment which gives $1/\mu_p$. Equation \eqref{eq:MAXCUT_env_condition} ultimately bounds how much each edge is counted in the sum over subgraphs.  
    
\begin{figure}
        \centering
        \includegraphics{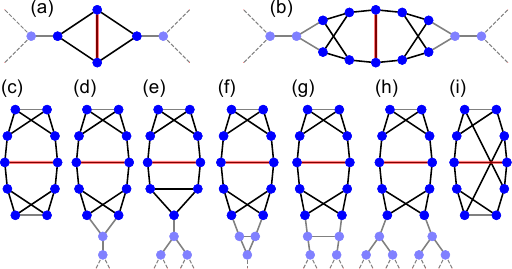}
        \caption{The worst case graph environments which bound the value of $\mu_p$ for (a) $p=1$ and (b) $p=2$, found by enumeration of all possible graph environments. These particular graph environments count the contribution of the central (red) uncut edge the maximal amount. There are 7 graph environments which have a larger sum than (b), shown as (c-i), which are excluded due to additional uncut edges.}
        \label{fig:worstcase_graph_env}
    \end{figure}

    For $p=1$ there are 6 possible graph environments (shown in Fig.~\ref{fig:subgraph1_graph_envs}). The worst-case graph environment is Fig.~\ref{fig:worstcase_graph_env}a, which is the central edge of two triangles. In this graph environment, the edge $\langle ij\rangle$ participates in an odd length (triangle) cycle in five subgraphs which each have five edges, and so Eq.~\eqref{eq:MAXCUT_env_condition} sums to 1. Thus, $\mu_1 = 1$.
    
    For $p=2$ there are a large number of graph environments. Like the $p=1$ case, the value $\mu_2$ can be found by searching through every possible combination of subgraphs an edge can participate in, e.g.~all graph environments, and finding the largest value of the sum. This search can be simplified by avoiding trivial instances of graph environments. If the edge $\langle ij\rangle$ does not participate in an odd length cycle in the subgraph where it is the center edge, it will not participate in an odd length cycle in any other subgraph of the graph environment. This excludes all subgraphs for which the center edge is the only connection between two sides, such as the tree graph, and so reduces the number of graph environments to a manageable amount. Under this exclusion, there are $1002191$ graph environments to search for $p=2$. 
    
    Evaluating the sum from Eq.~\eqref{eq:MAXCUT_env_condition} for each graph environment finds $117$ for which $\mu_2<1$. The worst case graph environment is shown in Fig.~\ref{fig:worstcase_graph_env}c. The central edge participates in an odd length cycle for all thirteen of the subgraphs it appears in; one subgraph has 13 edges, four have 12 edges, and eight have 9 edges, and so Eq.~\eqref{eq:MAXCUT_env_condition} gives $\mu_2$ as:
    
    \begin{equation}
        \frac{1}{13} + \frac{4}{12} + \frac{8}{9} = \frac{152}{117}=\frac{1}{\mu_2}.
    \end{equation}
    
    However, for this graph environment and the six following as ordered by Eq.~\eqref{eq:MAXCUT_env_condition} (Fig.~\ref{fig:worstcase_graph_env}c-i) we can evaluate the weighted sum on the right hand side of~\eqref{eq:uncut_lowerbound}  directly. We find that these graphs have at least one additional uncut edge in the \maxcut partition. For example, for subgraph environment \ref{fig:worstcase_graph_env}c, the center edge plus the two additional uncut edges give the weighted sum on the right hand side of~\eqref{eq:uncut_lowerbound} as $152/117 +95/126+95/126<3$, and similar for Fig.~\ref{fig:worstcase_graph_env}d-i.  Thus, these graph environments may be excluded to get a tighter bound on $R(\mathcal G)$.
    
    The eighth graph environment (Fig.~\ref{fig:worstcase_graph_env}b) has only one uncut edge in a \maxcut solution and so sets $\mu_2$. The central edge participates in an odd length cycle for all thirteen of the subgraphs it appears in; five of these subgraphs have 13 edges, and eight have 11 edges. Using Eq.~\eqref{eq:MAXCUT_env_condition}, this worst-case graph environment bounds $\mu_2$ to be $\mu_2=143/159$. Equation \eqref{eq:uncut_lowerbound_chain} thus simplifies to a lower bound on the number of uncut edges in a graph $\mathcal G$, using the fact that $R(\mathcal G_{\langle kl\rangle}^p)=\sum_{\langle ij\rangle}r_{\langle ij\rangle}^{\langle kl\rangle}$ and similar for $r_{\langle ij\rangle}$
    
    \begin{equation}
        \frac{143}{159}\sum_{\langle kl\rangle}\frac{R(\mathcal G_{\langle kl\rangle}^2)}{N(\mathcal G_{\langle kl\rangle}^2)}\leq R_M(\mathcal G).
    \end{equation}
    
    The constant $\mu_p$ serves as a guarantee that uncut edges in $\mathcal G$ are not over counted in the sum over subgraphs $\mathcal G_{\langle kl\rangle}^p$. Equivalently, the number of cut edges is a sum over subgraphs

    \begin{equation}\label{eq:maxcut_overestimate}
        C_M \leq \sum_\lambda N_\lambda(\mathcal G)c_\lambda\quad;\quad c_\lambda = 1-\frac{143}{159}\frac{R(S_\lambda)}{ N(S_\lambda)},
    \end{equation}
    where $c_\lambda$ is the local \texttt{MAXCUT} fraction for subgraph $\lambda$. Values for $c_\lambda$ are enumerated in Table \ref{tab:graph_menengere}. This combinatoric search emphasises a thesis of the paper: there could be some unexpected graph structure (as here) which in fact is a worst-case graph that can't be generated from simple intuition. Thus, any proof of performance guarantees must be combinatoric in nature.

    \subsection{Bounding the approximation ratio as a fraction of sums}
    
    The numerator $F_{M}$ is a lower bound and the denominator $C_M$ is an upper bound and so a fraction of sums serves as a lower bound on the approximation ratio
    
    \begin{equation}\label{eq:bounded_AR}
        C_p(\mathcal G)\geq \frac{\sum_\lambda N_\lambda(\mathcal G)f_\lambda}{\sum_\lambda N_\lambda(\mathcal G)c_\lambda}.
    \end{equation}
    
    Values for $f_\lambda$ and $c_\lambda$ for the enumerated subgraphs of $p=1,2$ and fixed degree $\nu=3$ are shown in Table \ref{tab:graph_menengere}, and details of the computation of $f_\lambda$ are shown in appendix \ref{app:computing_params}.

     As an example, consider the graph shown in Fig. \ref{fig:example_tiling}. Each edge is labeled by the index of the unique subgraph identified from $\mathcal G_{\langle ij\rangle}^p$. For $p=1$, there are 4 instances of subgraph 0 (``the tree"), 10 instances of subgraph 1 (``single triangle"), and 1 instance of subgraph 2 (``two triangles"). For $p=2$ there are 6 different kinds of subgraphs. Equation~\eqref{eq:bounded_AR} lower bounding the approximation ratio for $p=1$ becomes

    \begin{equation}C_1(\mathcal G) \geq \frac{4f_0 + 10f_{1} + f_2}{4c_0 + 10c_{1} + c_2}.
    \end{equation}

    Using Table \ref{tab:graph_menengere}, one can look up the expectation values and local \maxcut values. The upper bound on the \maxcut value is $12.8\geq 12$, the exact value, and the approximation ratio for this particular graph will be at least $C_1\geq 0.759$ and $C_2\geq 0.808$ for $p=1$ and $2$, respectively.

    In this way, a lower bound on the approximation ratio of any graph can be found. This lower bound is rather pessimistic, as it chooses seemingly arbitrary angles $(\gamma,\beta)$. However, the particular choice of angles given by Eq.~\eqref{eq:angle_choices} appear to still have large expectation values for all subgraphs. This fact will be discussed later.

    \section{worst case for 3-regular graphs}\label{sec:worst_case_p2}
    
     What is the worst case approximation ratio for a particular fixed value of $p$ and given set of graphs $\{\mathcal G\}$? There exists some graph $\mathcal G_*\in \{\mathcal G\}$ which can be chosen maliciously such that the maximal approximation ratio $C(\mathcal G_*)$ is minimal in $\{\mathcal G\}$. This graph $\mathcal G_*$ represents a problem instance for which a QAOA device with fixed $p$ has the worst performance. Any other graph will have a larger approximation ratio and thus this worst case is a performance guarantee on QAOA.
    
    Na\"ively, finding such a graph $\mathcal G_*$ is hard. The number of possible graphs is exponential in the number of vertices and we are interested in the general performance for arbitrarily large graphs, so a simple search through many graphs will not work. Because of this, a more careful approach must be taken to find lower bounds on the approximation ratio. Two methods are presented below.
    
    The first method obtains a lower bound by finding worst case combinations of subgraphs which may or may not form a consistent graph. By considering more and more complicated combinations of subgraphs, one can get a tighter bound from below on the minimum approximation ratio. This is the approach used for the original $p=1$ bound by Farhi et al.~\cite{farhi2014}.
    
    The second method presents a graph hierarchy which finds that the class of graphs with no cycles less then 4 (for $p=1$) or 6 (for $p=2$) are worst case. This is done by finding that, given a graph $\mathcal G$, there always exists a new graph $\mathcal G'$ with a smaller or equal approximation ratio. This is done by replacing edges with a subgraph to reduce the number of small cycles in the graph.

    \subsection{Lower bounds for $p=1$}\label{sec:p1_lowerbound}
    
    Instead of finding the exact approximation ratio of the worst case graph, one can instead \textit{bound} the worst case approximation ratio from below, by only including subgraphs with a small approximation ratio. This is an extension of the original analysis of Farhi et al.~\cite{farhi2014}. By Eq.~\eqref{eq:bounded_AR} a lower bound can be found by decomposing a graph $\mathcal G$ into subgraphs of a particular set $\{\mathcal S_\lambda\}$. Consider the inequality
    
    \begin{equation}\label{eq:subgraph_ordering}
        \frac{F}{C}\leq \frac{f'}{c'}\quad\Leftrightarrow\quad\frac{F}{C}\leq \frac{F+f'}{C+c'}\leq \frac{f'}{c'}
    \end{equation}
    for all $f,c,F,C>0$.  In context of Eq.~\eqref{eq:bounded_AR}, the terms are expectation values $F,f$ and local \maxcut values $C,c$ of two sets of subgraphs. One can order all subgraphs $S_\lambda$ by their own local approximation ratio $C_\lambda = f_\lambda/c_\lambda$ and constructively add subgraphs, starting with the subgraph with the smallest local approximation ratio.
    
    By Eq.~\eqref{eq:subgraph_ordering}, including only the worst subgraph of a graph, or excluding the best subgraph, gives a lower bound on the global approximation ratio. Including any other subgraph with a larger local approximation ratio will only increase the value, and excluding a subgraph with a larger local approximation ratio will only decrease the value.
    
    Taking Fig. \ref{fig:example_tiling} as an example graph, one can order the local approximation ratios $f_0/c_0\leq f_2/c_2\leq f_1/c_1$. Successive lower bounds on the approximation ratio of the graph can be found by including more and more subgraphs, eg
    
    \begin{equation}
        \frac{4f_0}{4c_0}\leq \frac{4f_0 + f_2}{4c_0+c_2} \leq \frac{4f_0 + 10f_{1} + f_2}{4c_0 + 10c_{1} + c_2} \leq C_1(\mathcal G).
    \end{equation}
    
    Adding additional subgraphs to the count gets a larger lower bound on the approximation ratio of the graph. If the graph $\mathcal G$ was worst case, then this ordering results in a strict lower bound on the approximation ratio.

    The worst case graph will include some number of each subgraph in its count $N_\lambda(\mathcal G_*)$. It is simple to iterate through the 3 possible subgraphs as enumerated in Table \ref{tab:graph_menengere} to find that subgraph 0, the tree, is minimal, with $f_0/c_0=0.692/1.000$. Thus, the approximation ratio for the worst case graph $\mathcal G_*$ is bounded from below by only including the minimal subgraph
    
    \begin{equation}
        C_1(\mathcal G)\geq C_\text{min}\geq0.692.
    \end{equation}
    
    The analysis for the minimum approximation ratio for the $p=1$ case ends here. This is because there are graphs which only include the tree subgraph, which have no cycles less than 4. The minimum approximation ratio $C_\text{min}\leq C(\mathcal G)$ for any graph $\mathcal G$ by definition; however, $C_\text{min}\leq C(\text{[tree]})=0.692$, and so the \textit{exact} minimum approximation ratio for $p=1$ is this value. This is the analysis of Farhi et al.~\cite{farhi2014}: they observe that the worst graph is made only of the tree subgraph, then observe that such a graph exists. This analysis does not hold for the $p=2$ case.
    
    \begin{figure}
        \centering
        \includegraphics[width=\linewidth]{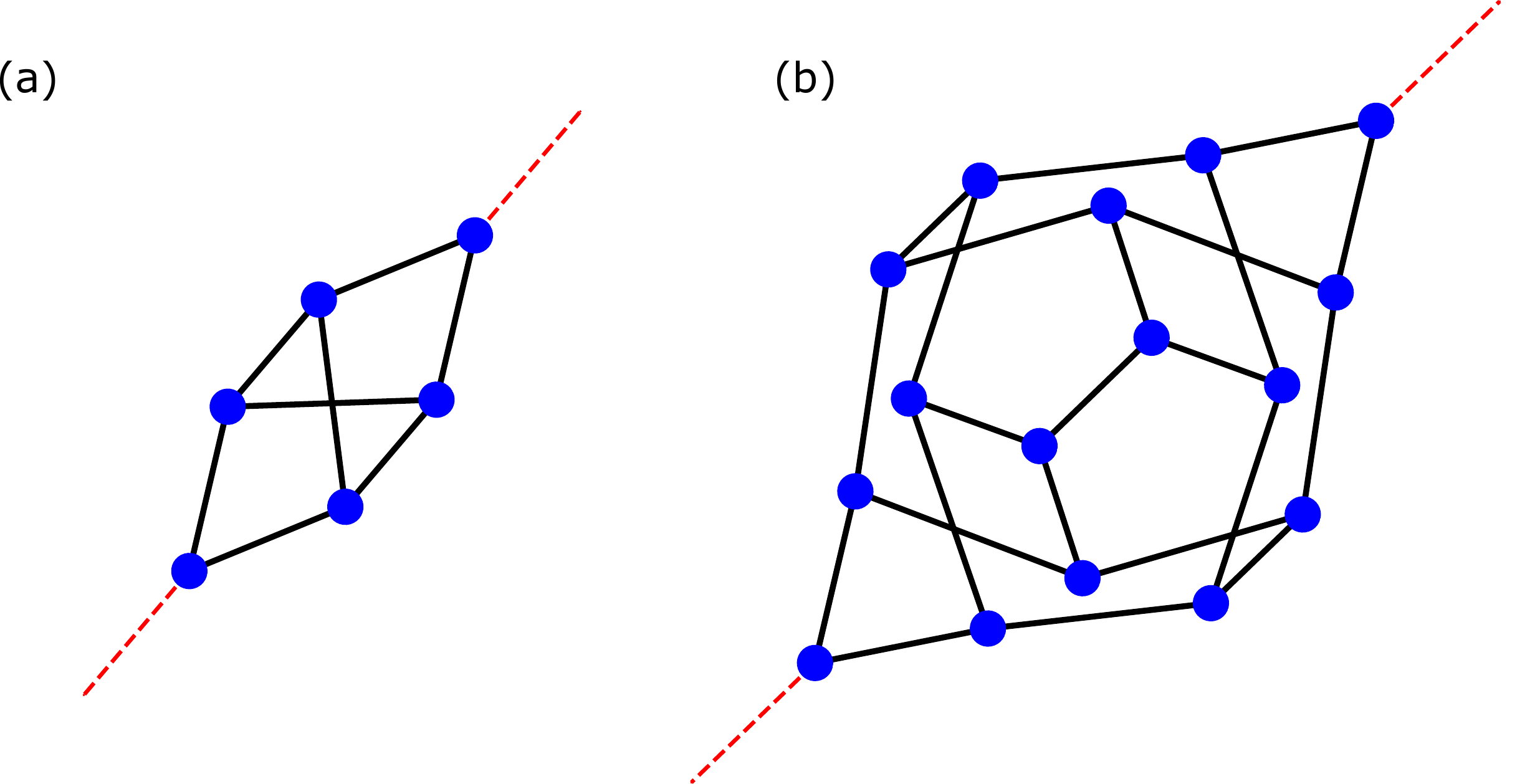}
        \caption{Edge replacements for $p=1$ and $p=2$ graphs. Given an edge $E$, the edge is replaced by this subgraph, where red dashed indicates the original edge. The left replacement has no cycles $\leq 3$, while the right has no cycles $\leq 5$, appropriate for $p=1$ and $p=2$, respectively.}
        \label{fig:cut_replacements}
    \end{figure}
    
    \subsection{Graph hierarchy for $p=1$}\label{sec:p1_hierarchy}

\begin{figure}
        \centering
        \includegraphics{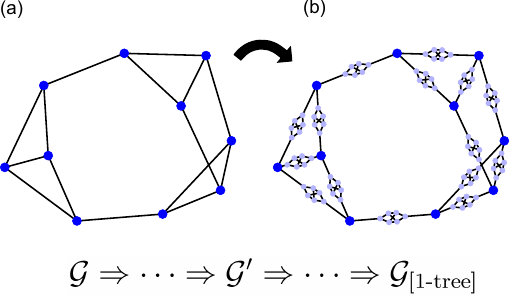}
        \caption{An example graph reduction. Each edge of some original graph $\mathcal G$ (a) is iteratively replaced with a 6 vertex subgraph, until eventually every edge is replaced to find a worst-case graph $\mathcal G_\text{[1-tree]}$ (b).}
        \label{fig:decorated_graph}
    \end{figure}

     Before continuing to the more difficult $p=2$ case, let us introduce a hierarchy of graphs for $p=1$ where, heuristically, graphs with fewer small cycles have a smaller approximation ratio. We will show that, given a graph $\mathcal G$, one can always find a new graph $\mathcal G'$ with $C_1(\mathcal G)\geq C_1(\mathcal G')$ unless the graph is of a specific class of graphs with no cycles of length $\leq 3$. We denote such graphs as ``1-tree graphs", which are constructed only out of the $p=1$ tree subgraph (See Fig.~\ref{fig:example_tiling} bottom left and Fig.~\ref{fig:decorated_graph}). Similar graphs can be defined for $p$-tree graphs, which have no cycles of length $\leq 2p+1$.
     
     Given a graph with small cycles, a new graph can be found with a worse approximation ratio, which proves inductively that the graph with no small cycles is the worst case graph via recursion. Given a worst case candidate graph $\mathcal G$ which is not a 1-tree, a 1-tree graph can be shown to have a smaller approximation ratio by recursion $\mathcal G\to \mathcal G'\to \mathcal G''\to \dots\to\mathcal G_\text{[1-tree]}=\mathcal G_*$. Thus, 1-tree graphs have a lower approximation ratio then any other graph. Let us continue by proving the graph reduction $\mathcal G\to\mathcal G'$.

    \quad
    
\begin{figure}
        \centering
        \includegraphics{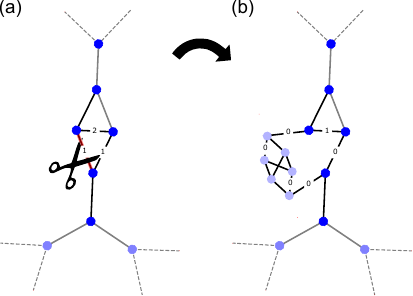}
        \caption{An example edge replacement operation. The red center edge in some graph environment (a) is replaced with 10 edges and 6 vertices (b), removing the original size-3 cycle. The new graph with the replaced edges will have a smaller approximation ratio. Two additional edges' subgraphs (labeled) are also modified.}
        \label{fig:example_cut}
    \end{figure}

     For a graph $\mathcal G$, choose an edge $E$ which participates in at least one cycle of length 3. Then, replace the edge with the 6-vertex subgraph of Fig.~\ref{fig:cut_replacements}a. This creates a new graph $\mathcal G'$, where the cycle of length 3 that the original edge participated in is replaced with a new cycle of length 7. An example is shown in Fig. \ref{fig:example_cut}. Let us prove that this new graph has a smaller approximation ratio.
    
    When this edge is replaced, the graph is modified and so the subgraphs  $\mathcal G_{\langle ij\rangle}^p$ of surrounding edges may also be modified. To prove the graph reduction, we must show that this replacement reduces the approximation ratio for all possible modifications of edges. All possible combinations of subgraph assignments to the edges of $\mathcal G_{\langle ij\rangle}^p$ is the set of graph environments of an edge, and so one may prove the graph reduction by checking some condition for all possible graph environments.

    A subset of graph environments are the relevant graph environments, which only include edges whose subgraph assignment is modified under replacement of the center edge. For $p=1$, there are 4 relevant graph environments, as replacing the central edge of Fig. \ref{fig:subgraph1_graph_envs}a,b,c does not change the subgraph assignment of its surroundings. Thus, the graph reduction will involve checking a condition for all possible relevant graph environments.

    Now, consider replacing a particular edge of some graph $\mathcal G$ with the subgraph of Fig. \ref{fig:cut_replacements}a yielding a graph $\mathcal G'$. The edges of the graph can be partitioned into two sets: edges which are replaced or have their subgraph assignment modified by the replacement procedure, which are found in the relevant graph environment, and edges which are not modified. The edges which are not modified have expectation values which sum to $F$ and local \maxcut sum $C$, and the edges which are replaced or modified have expectation value sum $f$ and local \maxcut sum $c$. The replaced or modified edges have a new expectation value sum $f'$ and local \maxcut sum $c'$, which include the additional 10 edges of the replacement operation. Now, consider the following clauses and their implication

    \begin{align}
        \mathcal A:\qquad&\frac{F}{C}\leq \frac{f}{c},\nonumber\\
        \mathcal B:\qquad&\frac{f}{c}\geq \frac{f'-f}{c'-c},\nonumber\\
        \mathcal C:\qquad&\frac{F + f}{C + c}\geq \frac{F + f'}{C + c'},\nonumber\\\nonumber\\
        &\mathcal A \wedge \mathcal B \Rightarrow \mathcal C.\label{eq:inequality_clauses}
    \end{align}
    
    Clause $\mathcal A$ is a restriction on choice of edge $E$. One must choose an edge such that, if the replaced or modified edges are removed from the count of subgraphs, the approximation ratio will decrease. By Eq.~\eqref{eq:subgraph_ordering}, this choice of edge will always exist by the ordering of subgraphs.
    
    Clause $\mathcal B$ is a condition on modifying subgraph assignments of an edge and its surroundings under replacement. This can be checked for every graph environment. For $c'-c>0$, which is the case when adding more edges under replacement, this clause is equivalent to $f/c\geq f'/c'$, that the new subgraph has a smaller local approximation ratio. Clause $\mathcal B$ generalizes to other graph modifications, such as reducing a 3 cycle to a single vertex.
    
    Clause $\mathcal C$ compares the lower bound approximation ratios for the graphs $\mathcal G$ and $\mathcal G'$ before and after a replacement procedure. The inequality states that the approximation ratio of the new graph will be smaller. If clause $\mathcal A$ and $\mathcal B$ are True, then the new graph has a smaller approximation ratio, which proves the graph reduction for a particular choice of edge replacement.
    
    As an example, consider Fig. \ref{fig:example_cut}, performing an edge replacement within some graph with the particular graph environment of Fig. \ref{fig:subgraph1_graph_envs}e. Here, there are 3 edges whose subgraph assignments will be replaced or modified by the replacement procedure, specifically the three edges of the triangle. Two edges have a subgraph assignment of 1 (single triangle), and one has subgraph assignment of 2 (two triangles). After the procedure, there are 11 copies of subgraph 0 (the tree), and one copy of subgraph 1 (an edge of a triangle): two from the original graph environment, plus an additional 10 from the cut replacement of Fig. \ref{fig:cut_replacements}. Using Table \ref{tab:graph_menengere}, one can compute
    
    \begin{equation}
        f=1.911;\quad  f'=8.253;\quad c=2.4;\quad c'= 11.8\nonumber
    \end{equation}
    
    It is simple to check that these values satisfy clause $\mathcal B$ of Eq.~\eqref{eq:inequality_clauses}. To prove that there always exists a graph reduction $\mathcal G\to\mathcal G'$, one can check all 4 of the relevant graph environments. It is found that clause $\mathcal B$ is true for all relevant environments. Thus, it is shown that, given a graph $\mathcal G$, a new graph $\mathcal G'$ can be constructed which will have a smaller or equal approximation ratio, done by replacing edges to remove cycles of length 3 in the graph. This creates a hierarchy of graphs for which graphs with fewer small cycles have a smaller approximation ratio
    
    \begin{equation}
        C(\mathcal G)\geq C(\mathcal G')\geq\dots\geq C(\mathcal G_\text{[1-tree]})=0.6924,
    \end{equation}
    here proved for the $p=1$ case and consistent with the lower bound and original results \cite{farhi2014}. An example graph reduction to a 1-tree is shown in Fig. \ref{fig:decorated_graph}. This hierarchy holds for the fixed angles of Eq.~\eqref{eq:angle_choices}, and so this performance guarantee holds for any graph evaluated at these fixed angles.
    
        \subsection{Lower bounds for $p=2$}

     Finding a lower bound for $p=2$ is more complicated, because each edge lives in a larger graph environment. These larger graph environments mean that the simple lower bound method for $p=1$ is no longer exact. This is because the subgraph assignment of an edge constrains the subgraph assignments of neighboring edges, as there are only a finite number of graph environments. As an example, consider Fig. \ref{fig:subgraph1_graph_envs}f, which is the only $p=1$ graph environment for the two triangle subgraph 2. Any graph which includes this subgraph must by necessity also include at least 4 instances of subgraph 1 (an edge of a triangle). One cannot construct a graph out of just subgraph 2 for $p=1$.
     
     More generally, even though one may have some count of subgraphs $N_\lambda$, there is no guarantee that there exists a graph which has that particular count $N_\lambda\neq N_\lambda(\mathcal G)$ $\forall \mathcal G$. A similar constraint holds for $p=2$: neighboring and next-nearest neighboring edges may be constrained to particular configurations of subgraph assignments, due to the finite numbers of graph environments.

\begin{figure}
        \centering
        \includegraphics{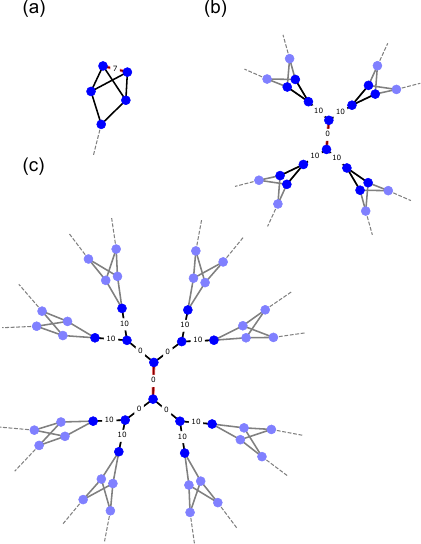}
        \caption{The three first worst case graph environments for $p=2$. For $N=1$ (a), the approximation ratio is $0.4886$. For $N\leq 5$ (b), it is $0.7376$. For $N\leq 14$ (c), it is $0.7424$. Each of these graph environments serves to find the lower bound on the approximation ratio of the worst case graph.}
        \label{fig:worstcase_k}
    \end{figure}

     The worst case graph is constructed out of some count of each kind of subgraph. A lower bound on its approximation ratio can be found by only including a certain subset of its subgraphs with a small local approximation ratio, even though the subset may not have an associated graph. As a first step, one can ignore this constraint and search through the 123 unique $p=2$ subgraphs of Table \ref{tab:graph_menengere} to find the subgraph with the smallest local approximation ratio. This is shown in Fig. \ref{fig:worstcase_k}a, with $f_{7}/c_{7}=0.4258/0.8571 = 0.4968$. By the argument of Section \ref{sec:p1_lowerbound}, adding any kind of any other subgraph will increase the approximation ratio, and so this number serves as a lower bound on the minimum approximation ratio for the $p=2$ case.
    
    However, as is clear from inspecting this subgraph, it is impossible to construct a graph out of only this subgraph. This means that any graph which includes this subgraph will also include some combination of other subgraphs, the inclusion of which increases the approximation ratio.
    
    Thus, this lower bound is loose, as no graph can be constructed with this approximation ratio, but any graph is guaranteed to have a larger approximation ratio. In fact, this bound is so loose that it is below the original $p=1$ bound, which still holds for $p=2$. This contrasts with the $p=1$ case, where there were graphs constructed out of only the worst case subgraph and the bound was tight.
    
    The next step in tightening this bound is to search through larger graph environments which also identify the subgraphs of the four neighboring edges to find a larger minimum approximation ratio. This graph environment is shown in Fig. \ref{fig:worstcase_k}b, and has an approximation ratio lower bounded by 0.7431, with two kinds of subgraphs. Again, it is impossible to construct a full graph out of just this graph environment: the outer edges are not allowed to be built of those two subgraphs, and thus any graph will have a strictly larger approximation ratio.
    
    Including graph environments one step larger identifies edges out to a depth 2 and finds the graph environment of Fig. \ref{fig:worstcase_k}c, with an approximation ratio of at least 0.7461. Beyond this limit, it becomes infeasible to find larger graph environments, due to the rapid growth in their number.
    
    Note that, unlike the simpler $p=1$ case, one cannot get an exact lower bound, and is instead recursively improvable by searching through larger and larger graph environments. In the next Section we consider how to approach the exact lower bound from above using the graph hierarchy.

    \subsection{Graph hierarchy for $p=2$}\label{sec:p2_tree_worstcase}

    Let us proceed by repeating the graph hierarchy argument of $p=1$ for the $p=2$ case. This case is complicated by having many more potential graph environments, because each subgraph is sensitive to a larger portion of its surroundings. Given a graph $\mathcal G$, a new graph $\mathcal G'$ can be found with $C_2(\mathcal G)\geq C_2(\mathcal G')$. As in the $p=1$ case, this is done by choosing a specific edge and replacing it with that of in Fig. \ref{fig:cut_replacements}b, which is a subgraph of 16 vertices and 25 edges.
    
    Similarly to the $p=1$ case, one can show that this graph reduction $\mathcal G\to\mathcal G'$ leads to a smaller approximation ratio by doing an edge replacement for every possible relevant graph environment, and checking the clauses of Eq.~\eqref{eq:inequality_clauses} for each. When an edge is replaced, edges up to two steps away from the replaced edge may have their subgraph assignments changed, as such subgraphs include all vertices within two steps of their center edge. Thus, one must check all $p=2$ graph environments. These can be found via the methods of Section \ref{app:finding_graphs}, finding all $p=4$ subgraphs subject to the constraint that there are no cycles $> 5$ in the minimum cycle basis \cite{Paton1969}.
    
    However, there are at least 30 billion $p=2$ graph environments, which is found by estimating a combinatorial lower bound on the number of graph environments for the $p=2$ tree subgraph. Instead, we find only the relevant graph environments. These relevent graphs are found by attempting to enumerate all relevant $p=4$ subgraphs in parallel starting with $p=3$ seed subgraphs, only including edges whose subgraph assignment is modified under center edge replacement. There are found to be 7058 such relevant graph environments; some examples are shown in Fig. \ref{fig:configs2}.
    
    The proof of graph reduction for $p=2$ is as follows. For each relevant $p=2$ graph environment, replace the special center edge and check the clauses $\mathcal A\wedge \mathcal B$ of Eq.~\eqref{eq:inequality_clauses}. We find that clause $\mathcal B$ is not satisfied for every relevant graph environment; however, there are no relevant graph environments for which $\mathcal A$ (the condition on choice of edge) is False and $\mathcal B$ is True. This puts a condition on choice of edge to be replaced: one must choose an edge such that $\mathcal A$ is True, by only choosing edges whose inclusion increases the approximation ratio, for which $\mathcal B$ will be True.

    This confirms the graph hierarchy for $p=2$. For each graph $\mathcal G$ with some cycles of length $\leq 5$, choose an edge and surrounding relevant graph environment $\mathcal G_{\langle ij\rangle}^{2p}$ which, upon removing it from the calculation of the approximation ratio, decreases the approximation ratio. This is forced by clause $\mathcal A$ of Eq.~\eqref{eq:inequality_clauses}. Upon replacing this edge with the 16-vertex edge replacement subgraph, the new graph $\mathcal G'$ will be guaranteed to have a smaller (or equal) approximation ratio.
    
    Inductively, this constructs a graph where every edge is replaced by the subgraph of Fig.~\ref{fig:cut_replacements} and has no cycles of length $\leq 5$, constructed only out of the tree subgraph. The expectation value of the tree subgraph and thus minimum approximation ratio is
    
    \begin{equation}
        C_2 \geq 0.7559.
    \end{equation}
    where worst case graphs are 2-trees, which have no cycles $\leq 5$. This is consistent with the observation in \cite{farhi2014}.

    In this Section, we have found worst case approximation ratios for $p=1$ and $2$ QAOA. Extending the original methods of \cite{farhi2014}, we find a recursively improvable lower bound of $C_2\geq 0.7424$, by considering larger and larger graph environments. Unlike the $p=1$ case, this lower bound cannot be made exact due to the adjacency restrictions implicit in the construction of graph environments. Using a recursive graph reduction and combinatoric proof, we find that 2-tree graphs, which have no cycles $\leq 5$, are exactly worst case. For every graph which is not a 2-tree graph, a new graph can be found with a smaller approximation ratio by finding some particular edge and replacing it with a 16 vertex subgraph which has no cycles $\leq 5$. Applied recursively, this eventually turns every edge of the original graph into one of these subgraphs, which is a worst case 2-tree graph.

    \section{Fixing variational parameters}\label{sec:angle_choice}
    
    We have shown that every graph $\mathcal G$ has an approximation ratio of $C_2\geq 0.7559$. However, this result is more general. The choice of angles $(\gamma,\beta)$, instead of being optimized for the particular graph, is fixed to a particular choice given by Eq.~\eqref{eq:angle_choices}. This means that this performance guarantee is stronger: For \textit{fixed} angles and any graph $\mathcal G$, the bound still holds
    
    \begin{align}
        &C_1\big(\mathcal G,\{35^\circ\},\{22^\circ\}\big)\geq 0.6924,\\
        &C_2\big(\mathcal G,\{28^\circ,31^\circ\},\{51^\circ,17^\circ\}\big)\geq 0.7559.
    \end{align}
    
    This particular set of angles is useful for experiments: using them with any graph guarantees a particular approximation ratio without the need of a classical optimizer back end. Additionally, we find numerically that gradient descent optimization from these angles finds the global optimum for almost every graph.
    
    It is clear why these angles were chosen: this set of angles is optimal for the worst case $p$-graphs. The minimum approximation ratio is found by minimizing over the set of all graphs while maximizing over angles. Any other choice of angles may have been valid, but may not have resulted in a tight minimum approximation ratio or even have the graph hierarchy be true. In fact, this particular choice of angles generates expectation values on subgraphs which are close to the global maximum of each subgraph, which can be seen comparing rows 3 and 5 in Table \ref{tab:graph_menengere}. There is no reason a priori for this to be so.
    
    \begin{table}[]
\begin{tabular}{|c|c|c|c|}
\hline
$\quad\quad\gamma_1\quad\quad$ & $\quad\quad\beta_1\quad\quad$ & $\quad\quad\gamma_2\quad\quad$ & $\quad\quad\beta_2\quad\quad$\\\hline
$35.3^\circ$ & $22.5^\circ$ & - & -\\ 
$144.7^\circ$ & $22.5^\circ$ & - & -\\ 
$215.3^\circ$ & $67.5^\circ$ & - & -\\ 
$324.7^\circ$ & $67.5^\circ$ & - & -\\
&&&\\
$28.0^\circ$ & $31.8^\circ$ & $51.4^\circ$ & $16.8^\circ$ \\ 
$28.0^\circ$ & $31.8^\circ$ & $231.4^\circ$ & $73.2^\circ$ \\ 
$152.0^\circ$ & $31.8^\circ$ & $128.6^\circ$ & $73.2^\circ$ \\ 
$152.0^\circ$ & $31.8^\circ$ & $308.6^\circ$ & $16.8^\circ$ \\ 
$208.0^\circ$ & $58.2^\circ$ & $51.4^\circ$ & $73.2^\circ$ \\ 
$208.0^\circ$ & $58.2^\circ$ & $231.4^\circ$ & $16.8^\circ$ \\ 
$332.0^\circ$ & $58.2^\circ$ & $128.6^\circ$ & $16.8^\circ$ \\ 
$332.0^\circ$ & $58.2^\circ$ & $308.6^\circ$ & $73.2^\circ$ \\ 
\hline
\end{tabular}
\caption{The anglar parameters for the 4 degenerate maxima of the $p=1$ tree subgraph, and 8 degenerate maxima of the $p=2$ tree subgraph. The expectation value of the objective function of any 3 regular graph evaluated at any of these angles is equal, and the approximation ratio is guaranteed to be above $0.6924$ and $0.7559$, respectively.}\label{tab:equivalent_angles}
\end{table}

    We also find that these angles are not unique. The landscape of expectation values $F(\gamma,\beta)$ is periodic modulo $2\pi$ in $\gamma$ and modulo $\pi/2$ in $\beta$. This is due to $SU(2)$ and $\mathbb Z_2$ symmetry (see Appendix \ref{app:computing_params} for details). Within these bounds, we find 4 degenerate maxima for $p=1$, and 8 degenerate maxima for $p=2$. The angles for the $p=1$ and $2$ tree subgraph are shown in Table \ref{tab:equivalent_angles}. Further, we find that for each subgraph, the expectation value of the center edge of that subgraph is the same evaluated at each of the 8 angles of any other subgraph.
    
    This means that for all graphs, the expectation value of the objective function will be the same for each of the 4 angular maxima of any subgraph; for example,
    
    \begin{equation}
        C_1(\mathcal G,35^\circ,113^\circ)=C_1(\mathcal G,145^\circ,23^\circ)\geq 0.6924
    \end{equation}
    for any graph and $p=1$, within numerical precision, and so forth for the additional 2 angles of Table \ref{tab:equivalent_angles}. Any of these angles provide a good starting point for an experimentalist needing good variational parameters.

    \section{Conjecture: $p$-Tree graphs are worst case}\label{sec:loop_conjecture}
    
    We found that replacing an edge with a subgraph with no small cycles results in a smaller approximation ratio for the $p=1,2$ cases. It is reasonable to expect that this behaviour should extend to larger $p$. This naturally leads to the following conjectures
    
    \quad
    
    \textbf{Graph hierarchy conjecture:} For any fixed $p$, graph $\mathcal G$, and fixed angles $\vec \gamma$, $\vec \beta$ optimal to the tree subgraph, there exists an edge replacement with a subgraph generalized from Fig. \ref{fig:cut_replacements} which results in a graph with a smaller approximation ratio.
    
    \quad
    
    In other words, there is a hierarchy of graphs, for which graphs with many small cycles less than $2p+2$  will have a better quality QAOA result than graphs with few cycles. This is shown to be true for $p=1$ and $2$ in Section \ref{sec:p1_hierarchy} and \ref{sec:p2_tree_worstcase}. For larger $p$ the edge replacement must be a larger subgraph with no cycles less than $2p+2$. This conjecture has two corollaries.
    
    \quad
    
    \textbf{Large loop conjecture:} The worst case graphs for fixed $p$ are $p$-trees, which have no cycles less than $2p+2$.
    
    \quad
    
    This conjecture is well motivated physically. When an edge is replaced, the algorithm ``sees" less of the full graph, due to the fact that it only knows of relations between vertices $\leq p$ steps away \cite{farhi2020}. This lesser knowledge of the full graph leads to worse answers, as the QAOA algorithm is then oblivious to improved solutions which ``see" more of the graph. Similarly, having no ``visible" cycles means the algorithm cannot distinguish between large cycles of even vs.~odd length, and thus cannot make good cut estimates which require this distinction. The worst case graph for all $p$ would be the Bethe lattice.
    
    \quad
    
    \textbf{Fixed angle conjecture:} Any graph evaluated at fixed angles optimum to the tree subgraph will have an approximation ratio larger than the guarantee.
    
    \quad
    
    Angles optimal to the tree subgraph for larger $p$ (e.g. an expansion of Table \ref{tab:equivalent_angles}) should result in \maxcut answers to any graph approximation ratios guaranteed to be above some value. These angles could be used as initial points for optimizers, or even excluding the optimization loop to compute good answers without feedback. The computation of these optimal angles for larger $p$ is left to future work. This conjecture is consistent with the phenomena of concentration \cite{brandao2018}, wherein optimal angles appear to be mostly independent of graph instance.
    
    \subsection{Worst case approximation ratio for $p=3$}
    
    Under the large loop conjecture, worst case graphs for $p=3$ are $3$-trees, with no cycles $\leq 7$ and constructed only out of the 30-vertex tree subgraph, which has no cycles. Using the same methods for the $p=1$ and $2$ case it is possible to compute the expectation value $f_0$ for this subgraph, which is thus the worst case performance guarantee for $p=3$ under the large loop conjecture. This subgraph has 30 vertices with a Hilbert space dimension of $2^{30}$. Using the symmetries of the tree subgraph, this can be reduced to a dimension of 1,631,721$\approx 2^{20.6}$, with an additional factor of 1/2 due to spin flip symmetry (see Appendix \ref{app:computing_params} for details). Optimization of angles using the methods of Appendix \ref{app:computing_params} finds
    
    \begin{equation}
        C_3\geq 0.7924
    \end{equation}
    under the large loop conjecture that 3-tree graphs are worst case for $p=3$. In principle, this bound can be made rigorous by searching through every possible $p=3$ graph environment and checking the inequalities of Eq.~\eqref{eq:inequality_clauses} for each. However there are 913,088 unique $p=3$ graphs and a much larger number of graph environments, which must extend up to 6 steps away from the replaced edge. While the combinatoric proof is in principle possible as the task is extremely parallelizable, we leave this challenging calculation to future work.
    
    \section{Comparison to classical algorithms}\label{sec:classical_compare}

    Here we compare performance bounds to the best classical algorithms. The most na\"ive classical algorithm is a random guess; it is simple to see that this cuts on average half the edges and so has an approximation ratio of at least $0.5$ \cite{Sahni1976}. It is known that calculating a cut with approximation ratio $\geq 16/17\approx 0.9412$ is NP-Hard \cite{Hrastad2001}. The algorithm of Goemans and Williamson \cite{goemans1995} gives an approximation ratio of at least $0.8786$ using semidefinite programming, and holds for any graph. For particular subsets of graphs this approximation ratio can be higher; for example, planar graphs can be solved efficiently in polynomial time \cite{Deza1994}. 3-regular graphs, which are the subject of this paper, have a lower bound of at least $0.9326$ \cite{Halperin2004}, also using semidefinite programming.
    
    Even comparing to the general Goemans-Williamson algorithm, $p=2$ QAOA does not achieve quantum advantage, as $0.7559<0.8786$. This does not discount the possibility that QAOA has better performance on a subset of graphs than any classical algorithm on the same subset. Finding such subsets of graphs is challenging for two reasons. First, for a particular subset of graphs there may exist some classical algorithm improving on the Goemans Williamson bound, as for planar graphs or 3-regular graphs, but finding such a specialized algorithm may be nontrivial. Second, the analysis of the subgraph structure of Table \ref{tab:graph_menengere} dictates the subset of allowed graphs. By the graph ordering conjecture, graphs with many cycles will have a better approximation ratios and thus have the most potential for finding instances with quantum advantage. Finding such subsets is beyond the scope of this paper.
    
    It is a curious fact that while the classical algorithms of this paper might be able to \textit{find} \texttt{MAXCUT} instances which have quantum advantage, it may not be possible to \textit{solve} those instances classically. This is because sampling bitstrings from a QAOA wavefunction is in complexity class \texttt{\#P}, and having such an algorithm would collapse the polynomial hierarchy at the third level \cite{farhi2016sampling}.
    
    \section{Upper bounds on minimum approximation ratio}\label{sec:upper_bounds}
    
    \begin{figure}
        \centering
        \includegraphics{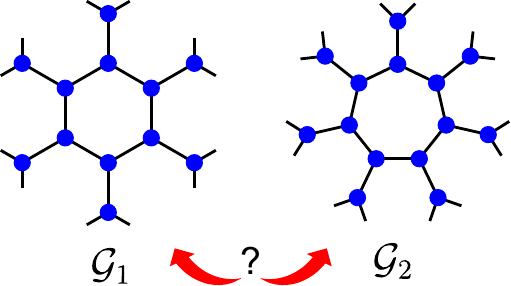}
        \caption{$p=2$ QAOA cannot distinguish between a tiling of hexagons ($\mathcal G_1$) or heptagons ($\mathcal G_2$), as both are constructed only from the tree subgraph. Graph $\mathcal G_1$ has a partition which cuts every edge, while graph $\mathcal G_2$ has at least one uncut edge per heptagon due to the odd length cycles. This puts upper bounds on the expectation value of the tree graph, and thus the minimum approximation ratio. This generalizes for all $p$ and graph connectivity $\nu$.}
        \label{fig:indistinguishable_graphs}
    \end{figure}
    
    While computing the minimum approximation ratio for $p>3$ is challenging, it is reasonable to ask how the minimum approximation ratio behaves with $p$. As $p\to \infty$, the approximation ratio approaches 1 in accordance with the adiabatic theorem \cite{farhi2014}. How does it do so? One way to compute this behavior is to bound the minimum approximation ratio from above: for a particular $p$, it can be \textit{at most} some value.
    
    One way of finding such a bound is to consider pairs of graphs which are indistinguishable under some fixed $p$ QAOA. First, construct a bipartite graph $\mathcal G_1$ as a tiling of $q$-gons with cycles of length $q = 2p+2$. For example, for $p=1$ this is a square ladder, while for $p=2$ this is a hexagonal honeycomb lattice (see Fig. \ref{fig:indistinguishable_graphs}). Because all cycles are of even length, it is simple to find a partition which cuts every edge, so that $C_\text{max}=n_\text{edges}$.
    
    Next, construct a graph $\mathcal G_2$, which is a tiling of $q$-gons with cycles of length $q=2p+3$. As an example, for $p=1$ or $2$ this can be seen as pentagons or heptagons (non-metrically) tiled on some curved surface (see Fig. \ref{fig:indistinguishable_graphs}). For $N$ $q$-gons, there are $Nq/2$ edges. Because the cycles are of odd length, at least one edge per $q$-gon must remain uncut. Each edge is shared by two $q$-gons and so for $N$ $q$-gons there are at least $N/2$ uncut edges and $C_\text{max} = N(2p+2)/2$ cut edges.
    
    For both graphs, there are no cycles of length $\leq 2p+1$, which means only the tree subgraph contributes to the expectation value. Critically, the tree graph cannot distinguish between the two graphs, even though they have different \maxcut values. Consider
    
    \begin{equation}\label{eq:G1_approxratio}
        C(\mathcal G_1) =\frac{n_\text{edges} f_0}{C_\text{max}}=f_0,
    \end{equation}
    the approximation ratio of the bipartite graph $\mathcal G_1$ is simply the expectation value of the tree graph, as every edge is cut. By definition, $C_\text{min}\leq C(\mathcal G_1)$. One may also compute the approximation ratio of $\mathcal G_2$, which is bounded from above by 1.
    
    \begin{equation}
        C(\mathcal G_2) = \frac{n_\text{edges} f_0}{C_\text{max}}=\frac{(2p+3)f_0}{(2p+2)}\leq 1.
    \end{equation}
    
    This bound from above puts an upper bound on the expectation value $f_0$. In combination, Eq.~\eqref{eq:G1_approxratio} gives an upper bound on the minimum approximation ratio
    
    \begin{equation}\label{eq:Cmin_upperbound}
        C_\text{min}\leq C(\mathcal G_1)= f_0\leq \frac{2p+2}{2p+3}.
    \end{equation}
    
    For $p$ large the bound goes as
    
    \begin{equation}
        C_\text{min}\leq 1-\frac{1}{2p}.
    \end{equation}
    
    This bound is independent of the graph connectivity and is consistent with the convergence of 2-regular graphs observed in \cite{farhi2014}.
    
\begin{figure}
        \centering
        \includegraphics{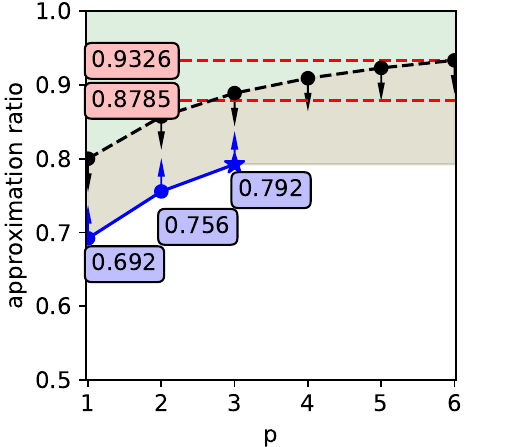}
        \caption{Results of the paper: approximation ratios vs. $p$. Blue line is the worst case approximation ratio for $p=1,2,3$; the $p=3$ case (star) assumes the large loop conjecture. Red dashed are the Goemans-Williamson \cite{goemans1995} and 3-regular \cite{Halperin2004} bounds. The minimum worst case approximation ratio is guaranteed to be below the black dashed line, which converges as $1-1/2p$.}\label{fig:goldenplot}
    \end{figure}

    Given a particular malicious \maxcut instance with no small cycles, the convergence of QAOA will be inverse polynomial with a power of \textit{at most} 1. Note that other graphs may converge much faster, as this bound only holds for graphs which have no cycles $\leq 2p+1$, which are exponentially large in $p$. This means for a fixed cycle length, after some very large $p$ convergence can be faster than this bound; the $p\to\infty$ behavior occurs only for the Bethe lattice, which has an infinite cycle length. The upper bound of Eq.~\eqref{eq:Cmin_upperbound} is plotted in Fig. \ref{fig:goldenplot}. The computed values for 1, 2, and 3 do not come close to this limit, as this bound is loose. From Section \ref{sec:classical_compare}, the best classical algorithms for 3-regular \texttt{MAXCUT} have an approximation ratio of at least $0.9326$; thus, $p$ must be at least greater than 5 to have quantum advantage for a general 3-regular graph. This argument is based on a particular graph feature and the $p\leq 3$ guarantees tend below the bound, and so one might have a more pessimistic estimate on $p$ due to special purpose classical algorithms and performance guarantees which may not saturate the bounds.
    
    \section{Conclusion}\label{sec:conclusion}
    
    Bounding the performance of near-term quantum algorithms is critical to understand where, how, and why quantum computers may gain advantage in the NISQ era. In this paper, we find a worst case performance guarantee for $p=2$ QAOA solving \maxcut on 3-regular graphs. This performance guarantee was found to be $C_2\geq 0.7559$, which holds for any graph evaluated not just at its optimized angles, but for a fixed set of angles given by Table \ref{tab:equivalent_angles}. Because this set of angles is fixed, they can act as good initial guesses for a classical optimizer with a guaranteed approximation ratio.

    More important than the number itself, the methods and the particular worst case graphs for which the bound was derived may be able to motivate particular ensembles of graph instances for which QAOA exhibits quantum advantage.
    
    The worst case graphs for $p=1$ and $2$ were proved to be graphs with no cycles $\leq 2p+1$. This was done via a graph reduction, replacing an edge with an expanded subgraph which distances the two original vertices of the edge and removes small cycles. The QAOA algorithm can only ``see" the structure of a graph within some small number of steps, and so the effect of the graph reduction is twofold. In removing and lengthening cycles, due to its local nature the algorithm cannot distinguish between large even and odd length cycles. In expanding the edge into a larger subgraph, the two previously adjacent vertices of the edge are distanced so their previous relation is obscured to the algorithm.
    
    These two properties stemming from the graph reduction suggest which graphs may have good QAOA solutions. Good graphs should have many small cycles, and should have a small-world structure \cite{Watts1998} where only a logrithmic number of steps is necessary to move from one vertex to any other.
    
    The properties presented here are heuristic and stem from the graph hierarchy, which is proved for $p\leq 2$ but can only be conjectured for $p>2$. Future work remains to find more rigorous specifications and properties of graphs and problem instances on which it may be possible to demonstrate quantum advantage in QAOA.
    
    \subsection*{Acknowledgements}
    This material is based upon work supported by the Defense Advanced Research Projects Agency (DARPA) under Contract No. HR001120C0068. We also thank Eddie Farhi and Sam Gutmann for their comments on \maxcut bounds.
    
    \normalem
	\bibliographystyle{apsrev4-1}
	\bibliography{citationlist} 
	
    \pagebreak
    
    \quad
    
    
    \appendix

    \section{Computing optimal parameters $\gamma,\beta$}\label{app:computing_params}
    
    This Appendix details computing the optimal objective function $f_\lambda$ for particular subgraphs. For each subgraph $\mathcal S_\lambda\in\{\mathcal S\}$, we wish to compute $\texttt{MAX}:$ $f_\lambda(\gamma,\beta)$. First, generate the local objective function for the subgraph by including one clause $\frac{1}{2}(1-\hat \sigma_z^i\hat \sigma_z^j)$ per edge $\langle ij\rangle$ in the subgraph, and similar for the local objective function on the special edge $\langle 0,1\rangle$ and mixing Hamiltonain $\hat B$ of $\hat \sigma_x^i$ for each vertex $i$ in the subgraph. We can compute expectation values exactly, which similarly enables access to derivatives of the objective function
    
    \begin{align}
        \partial_{\gamma_1} \langle \hat C\rangle& = \big[\partial_{\gamma_1}\langle \gamma,\beta|\big] \hat C |\gamma,\beta\rangle + \langle \gamma,\beta| \hat C \big[\partial_{\gamma_1}|\gamma,\beta\rangle\big]\\
        \partial_{\gamma_1}|\gamma,\beta\rangle&=-ie^{-i\beta_p \hat B}e^{-i\gamma_p \hat C}(\dots)e^{-i\beta_1 \hat B}\big(\hat C e^{i\gamma_1 \hat C}\big)|\gamma,\beta\rangle\nonumber
    \end{align}
    and similar for the other $\gamma,\beta$, with ellipsis denoting the other $2p-4$ generators. With access to both the exact expectation values and derivatives, parameters were optimized via a multistart gradient ascent algorithm. For each subgraph, initial parameters were chosen uniformly in parameter space, which is is compact in $\{[-\pi/4,\pi/4),[-\pi,\pi)\}^p$. Note that unlike for general QAOA, $\beta$ is periodic modulo $\pi/2$, due to $\mathbb Z_2$ symmetry, eg $\hat\sigma_z\to-\hat\sigma_z$. This is because the unitary over the mixing term
    
    \begin{equation}
        e^{i(\beta + \pi/2)\hat B} =\big(\prod_ie^{i\pi/2\hat\sigma_x^i}\big)e^{i\beta \hat B}=\hat{\mathbb Z}_2e^{i\beta \hat B},
    \end{equation}
    
    and $[\hat{\mathbb Z}_2,\hat B] = [\hat{\mathbb Z}_2,\hat C] = 0$. For each optimization, 25 random initial parameters were chosen to find a maxima with high probability. To find all degenerate maxima of a subgraph, initial parameters were chosen on a mesh, and each also optimized with gradient descent. At each step, the parameters are updated to change along the direction with the largest gradient, with size $0.075|\vec\nabla \langle C\rangle|$, where the constant is an implicit choice of the maximum second derivative. A particular optimization ends when the expectation value changes by less than $10^{-5}$.
    
    For the $p=3$ tree, a free optimization was used to speed up the calculation by reducing the Hilbert space using symmetry. Every graph has a certain set of isomorphisms, eg relabeling of each vertex. These isomorphisms define swap operators which commute with the generators $\hat B$ and $\hat C$, and thus the eigenvalues are good quantum numbers defining conserved subspaces of the Hilbert space. The initial wavefunction $|+\rangle$ is unchanged under any relabeling of indices, and so lives in the $+1$ subspace of all isomorphisms. While this reduction is in principle possible for any subgraph, it is particularly simple for the recognizable swap symmetries of the tree. Swapping any two ``branches" of the tree leaves the wavefunction invariant, and so only symmetric combinations remain. For two vertices (eg $p=0$), this is only three states: both up, both down, and the triplet state. For 6 vertices, there are 3 isomorphisms: swapping the left or right two vertices, or reflecting the three left with three right. Under these symmetries there are $21$ states. This can be done recursively by knowing that given two indistinguishable Hilbert spaces of dimension $\mathcal D$, there are only $\mathcal D(\mathcal D+1)/2$ symmetric combinations. Using this, for $p=2$ there are $903$ states; for $p=3$ there are 1,631,721 $\approx 2^{20.6}$; and for $p=4$ there are 5,325,028,475,403 $\approx 2^{42.3}$. There is an additional factor of $\approx 2$, as the generators have a $\mathbb{Z}_2$ spin flip symmetry $\hat \sigma_z\leftrightarrow -\hat \sigma_z$. In principle, it may be possible to simulate the worst case 4-tree exactly, as 41 qubits is within range of classical simulatability, but is beyond the scope of this work. It is reasonable to expect that these tree subgraphs can be simulated with tensor methods \cite{streif2019} up to $p\sim 10$.

        \section{Subgraph generation for fixed p}\label{app:finding_graphs}
    
    \begin{figure}
        \centering
        \includegraphics[width=\linewidth]{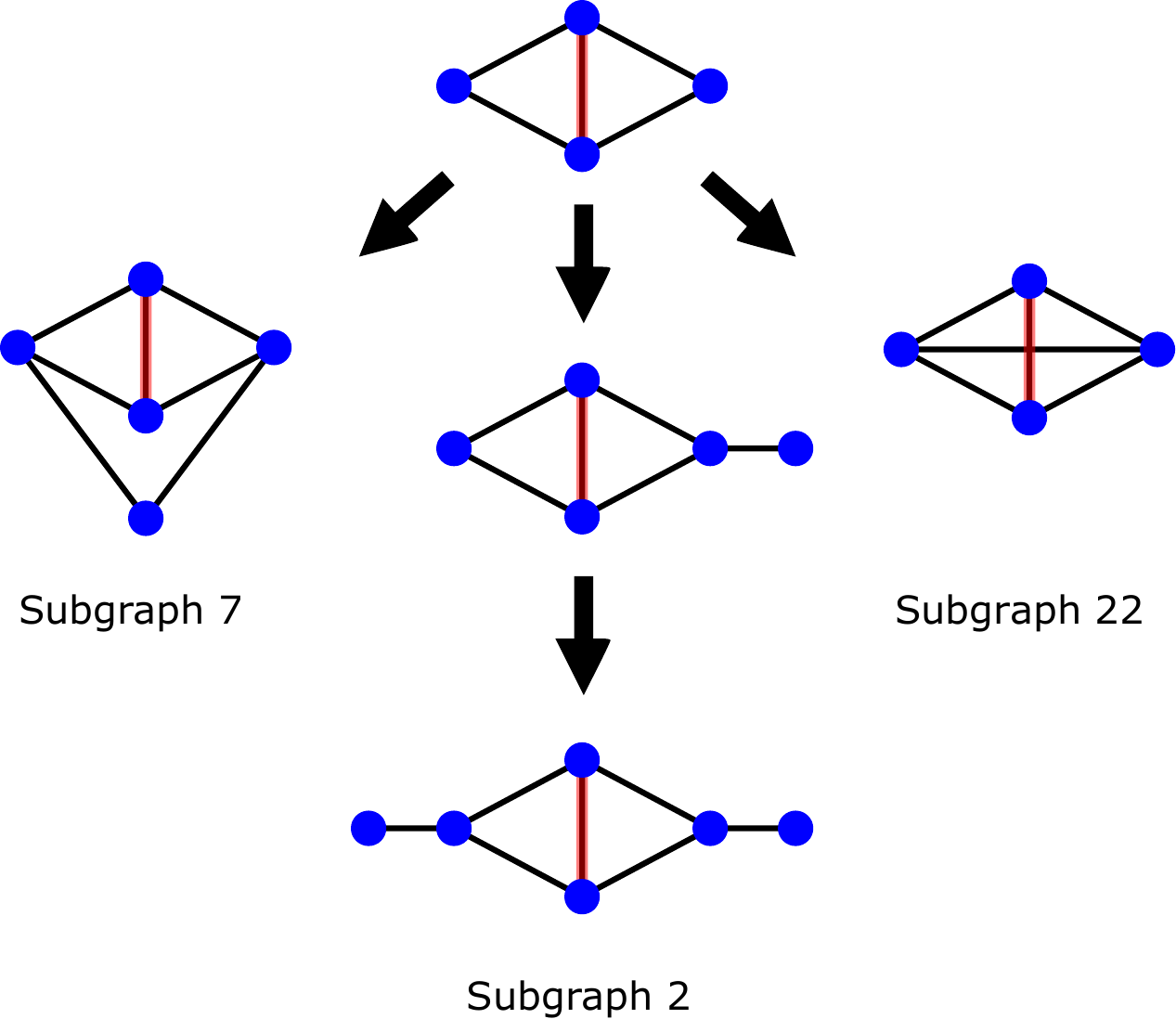}
        \caption{Recursively constructing some $p=2$ subgraphs. Starting from some seed $p=1$ subgraph (top), edges are added (arrows) to vertices which do not have 3 edges already, connecting to new or existing edges. Recursion through all possibilities finds all subgraphs, up to isomorphisms. The other two $p=1$ subgraphs have larger recursive trees, eventually finding all 123 $p=2$ subgraphs.}
        \label{fig:subgraph_recursion}
    \end{figure}
    
    In order to efficiently compute expectation values of graphs, one must go ``in reverse" to find all possible subgraphs $\mathcal G_{\langle ij\rangle}^p$. This appendix details the enumeration of the set of these subgraphs, denoted as $\{\mathcal S_\lambda\}_p$. First, find the set of all subgraphs $\{\mathcal S_\lambda\}_{p-1}$. For example, for $p=1$ this is the two-vertex graph connected by an edge; for $p=2$ these are the three $p=1$ subgraphs, and so forth, generated recursively. Next, for each of these subgraphs, find all the exterior vertices which have less than 3 edges (see Fig. \ref{fig:subgraph_recursion}). Then, iterate through adding one, two, or three edges. One may add one edge connecting two vertices of the original seed subgraph (Fig. \ref{fig:subgraph_recursion} Right), or add one edge connecting a vertex to a new vertex (Fig. \ref{fig:subgraph_recursion} Middle). Additionally, one may add two edges connecting two vertices to a new vertex (Fig. \ref{fig:subgraph_recursion} Left) or three edges connecting three vertices to a new vertex (not shown by Fig. \ref{fig:subgraph_recursion}). Iterating by adding these graphs to a heap, additional edges are added until all exterior vertices of the original subgraphs have three edges, and the heap is empty. This is guaranteed to find all subgraphs, as it searches through every possible permutation of new vertices connected to every combination of exterior vertices of the seed subgraph. When constructing all unique subgraphs, if a subgraph was isomorphic to an already-discovered graph, it is excluded from the heap.
    
    Using this recursion, there were found to be 3 subgraphs for $p=1$, 123 subgraphs for $p=2$, and $913,088$ subgraphs for $p=3$. The $p=1$ and $p=2$ subgraphs are enumerated in Table \ref{tab:graph_menengere}.

\begin{table}[]
\begin{tabular}{|c|c|c|}
\hline
\multirow{6}{*}{\includegraphics[height=55pt]{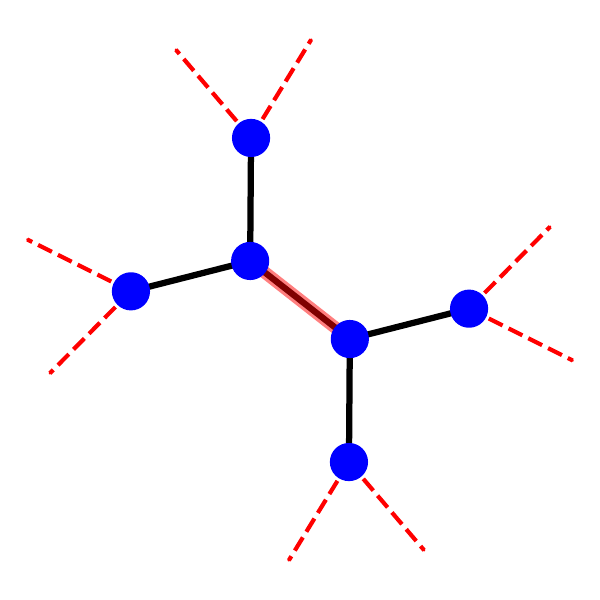}}
                & Index    & 1, 0 \\ \cline{2-3} 
                & $c_0$    & 1.000 \\ \cline{2-3} 
                & $f_{0}$    & $0.6924$ \\ \cline{2-3} 
                & Envs.    & 1 \\ \cline{2-3} 
                & $\gamma_{1}$    & $35^\circ$\\ \cline{2-3} 
                & $\beta_{1}$    & $22^\circ$ \\ \cline{2-3} 
            \hline 
\multirow{7}{*}{\includegraphics[height=55pt]{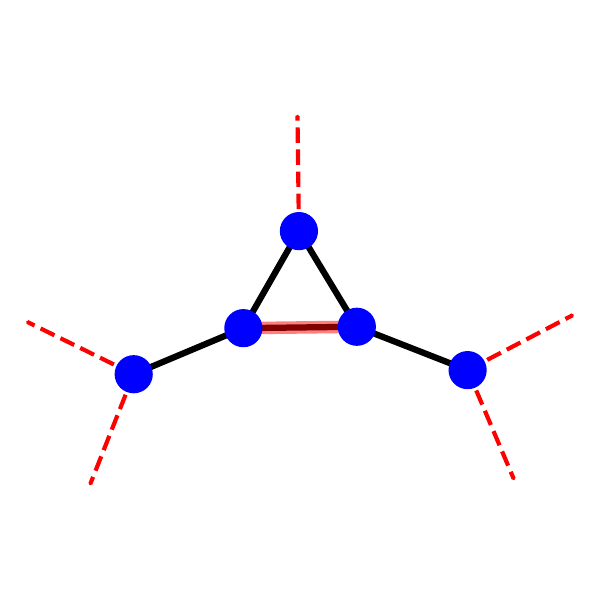}}
                & Index    & 1, 1 \\ \cline{2-3} 
                & $c_1$    & 0.8000 \\ \cline{2-3} 
                & $f_{1}$    & $0.6369$ \\ \cline{2-3} 
                & Envs.    & 2 \\ \cline{2-3} 
                & $f_{1}'$    & $0.6467$ \\ \cline{2-3}
                & $\gamma_{1}$    & $-211^\circ$\\ \cline{2-3} 
                & $\beta_{1}$    & $108^\circ$ \\ \cline{2-3} 
                \hline 
\multirow{7}{*}{\includegraphics[height=55pt]{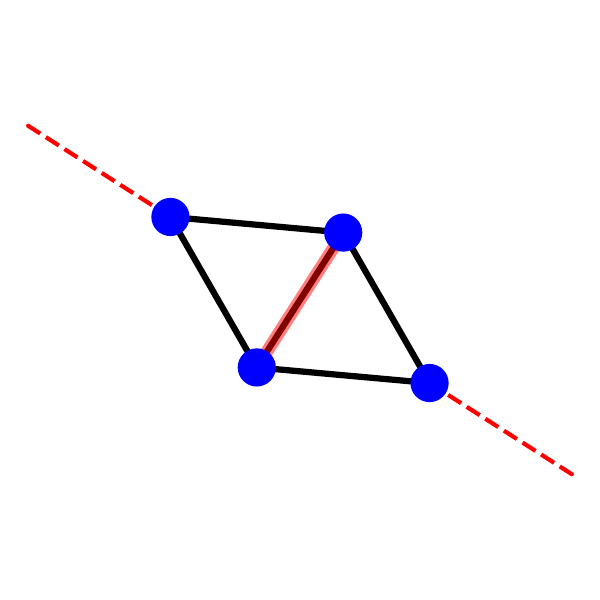}}
                & Index    & 1, 2 \\ \cline{2-3} 
                & $c_1$    & 0.8000 \\ \cline{2-3} 
                & $f_{2}$    & 0.5813 \\ \cline{2-3} 
                & Envs.    & 1 \\ \cline{2-3} 
                & $f_{2}'$    & 0.6163 \\ \cline{2-3} 
                & $\gamma_{1}$    & $-28^\circ$\\ \cline{2-3} 
                & $\beta_{1}$    & $164^\circ$ \\ \cline{2-3} 
                \hline 
\multirow{9}{*}{\includegraphics[height=100pt]{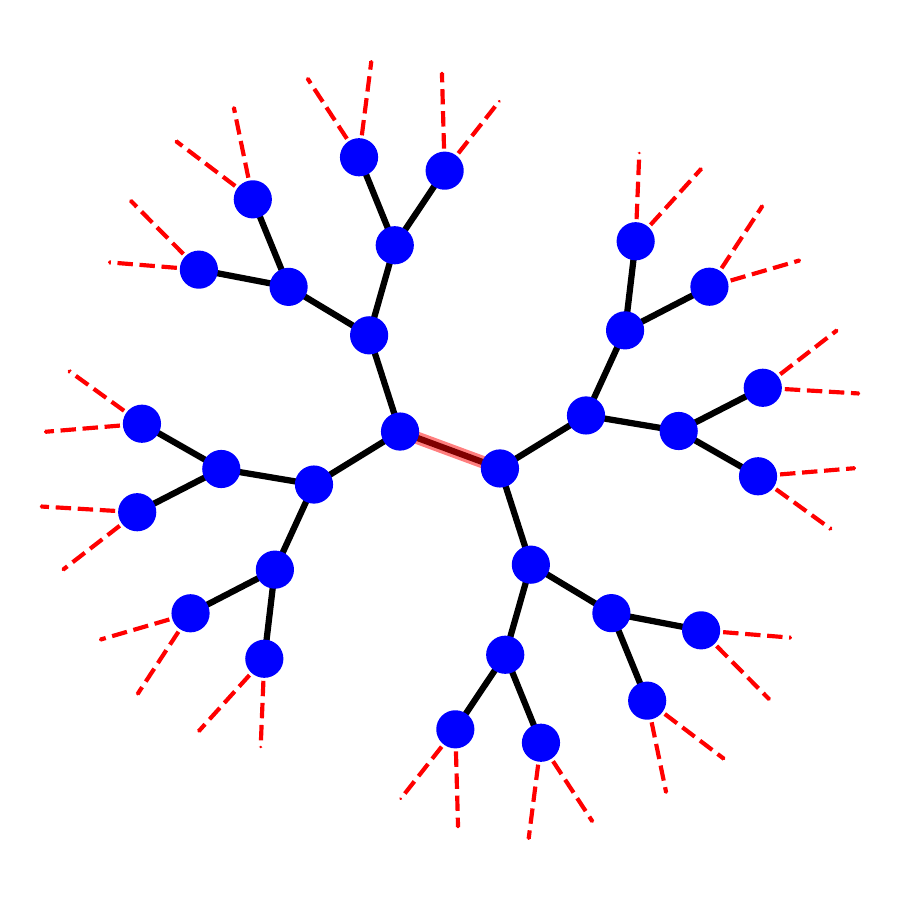}}
                & Index    & 3, 0 \\ \cline{2-3} 
                & $c_0$    & 1.000 \\ \cline{2-3} 
                & $f_{0}$    & 0.7924 \\ \cline{2-3}
                & Envs.    & 1 \\ \cline{2-3} 
                & $\gamma_{1}$    & $156^\circ$\\ \cline{2-3} 
                & $\beta_{1}$    & $-35^\circ$ \\ \cline{2-3} 
                & $\gamma_{2}$    & $-46^\circ$ \\ \cline{2-3} 
                & $\beta_{2}$    & $-27^\circ$ \\ \cline{2-3}
                & $\gamma_{3}$    & $-54^\circ$ \\ \cline{2-3} 
                & $\beta_{3}$    & $-14^\circ$ \\ \cline{2-3}\hline 
\end{tabular}
\caption{Numerical data for $p=1,2$ and  specific $p=3$ subgraphs. Row 1 enumerates subgraphs. Row 2 counts the local \texttt{MAXCUT} value, cutting N of M total edges. Row 3 is the expectation value for the edge evaluated at the angles of table \ref{tab:equivalent_angles}. Row 4 is the number of relevant graph environments which the subgraph can be in. Row 5 is the maximum expectation value of the subgraph, optimized at the angles of rows 6+. Image is the representation of the subgraph; red dashed represent edges connecting to the rest of the graph, while red solid represents the special center edge.}\label{tab:graph_menengere}
\end{table}


\include{graph_menengerie_p2_v8}

\end{document}

%% file: graph_menengerie_p2_v8.tex
\begin{table}[]
\begin{tabular}{|c|c|c|}
\hline
\multirow{8}{*}{\includegraphics[height=90pt]{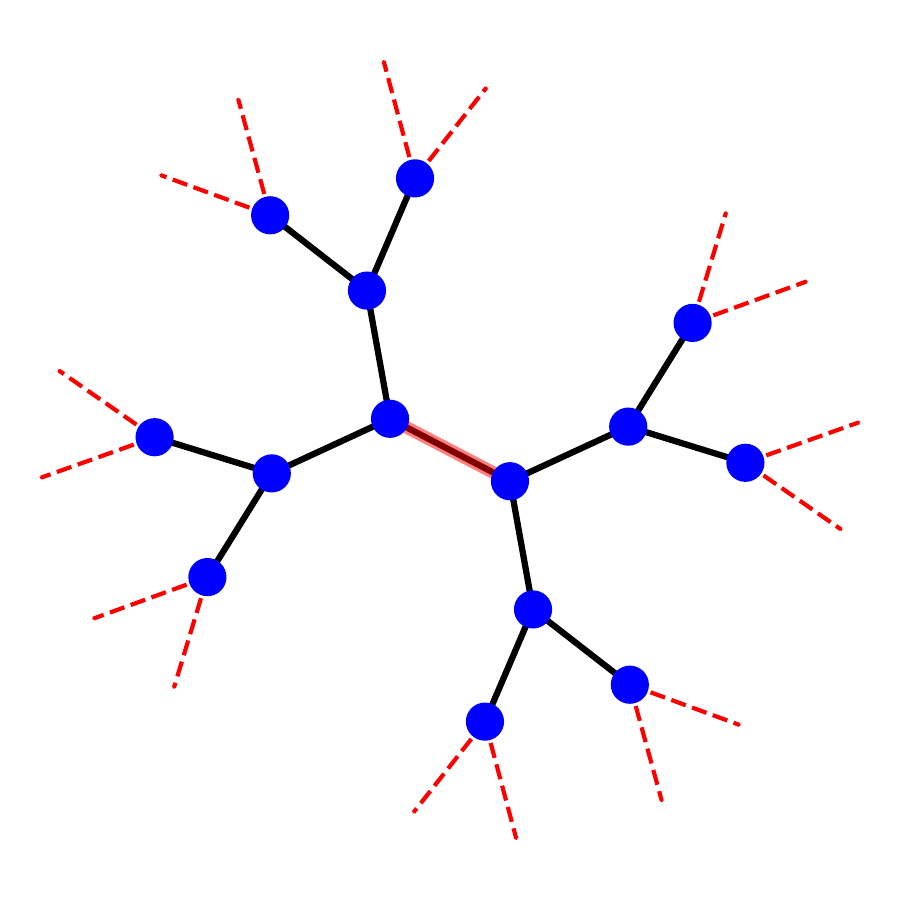}}
                & Index    & 2, 0 \\ \cline{2-3} 
                & $c_{0}$    & 1.0000 \\ \cline{2-3} 
                & $f_{0}$    & 0.7559 \\ \cline{2-3} 
                & Envs.    & 1 \\ \cline{2-3} 
                & $f_{0}'$    & 0.7559 \\ \cline{2-3} 
                & $\gamma_{1}$    & $28^\circ$\\ \cline{2-3} 
                & $\beta_{1}$    & $58^\circ$ \\ \cline{2-3} 
                & $\gamma_{2}$    & $52^\circ$ \\
\hline 
\end{tabular}
\end{table}
\begin{table}[]
\begin{tabular}{|c|c|c|}
\hline
\multirow{8}{*}{\includegraphics[height=90pt]{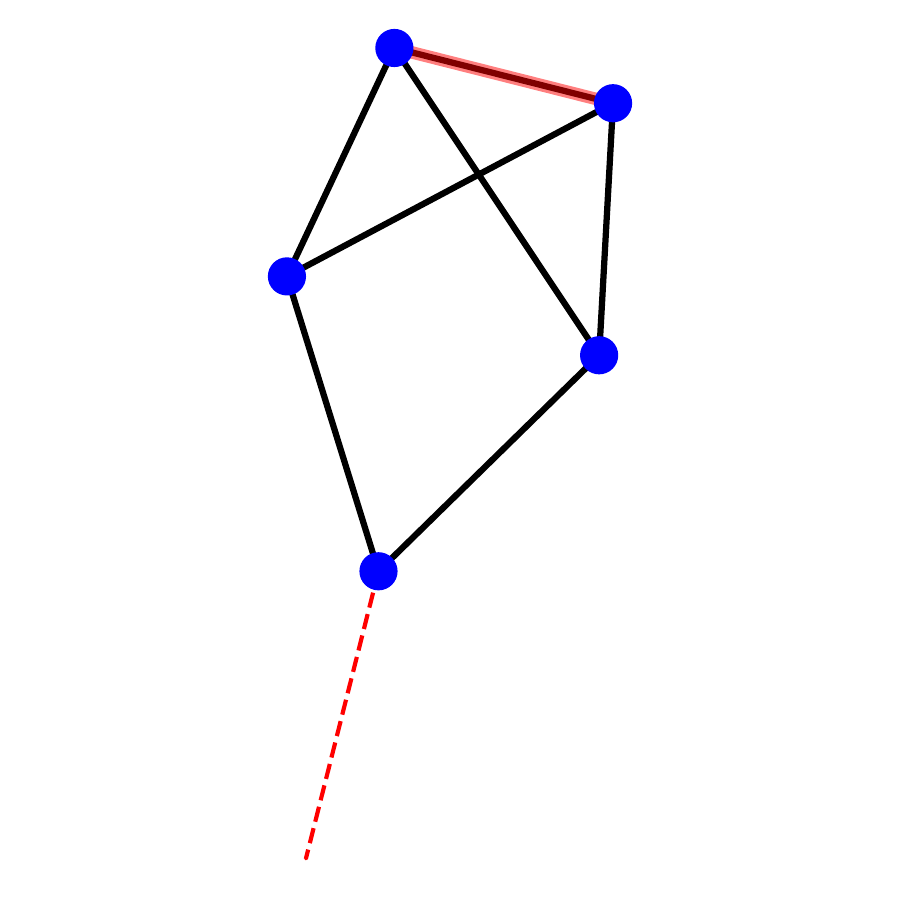}}
                & Index    & 2, 7 \\ \cline{2-3} 
                & $c_{7}$    & 0.8715 \\ \cline{2-3} 
                & $f_{7}$    & 0.4258 \\ \cline{2-3} 
                & Envs.    & 1 \\ \cline{2-3} 
                & $f_{7}'$    & 0.7492 \\ \cline{2-3} 
                & $\gamma_{1}$    & $217^\circ$\\ \cline{2-3} 
                & $\beta_{1}$    & $42^\circ$ \\ \cline{2-3} 
                & $\gamma_{2}$    & $107^\circ$ \\
\hline 
\end{tabular}
\end{table}
\begin{table}[]
\begin{tabular}{|c|c|c|}
\hline
\multirow{8}{*}{\includegraphics[height=90pt]{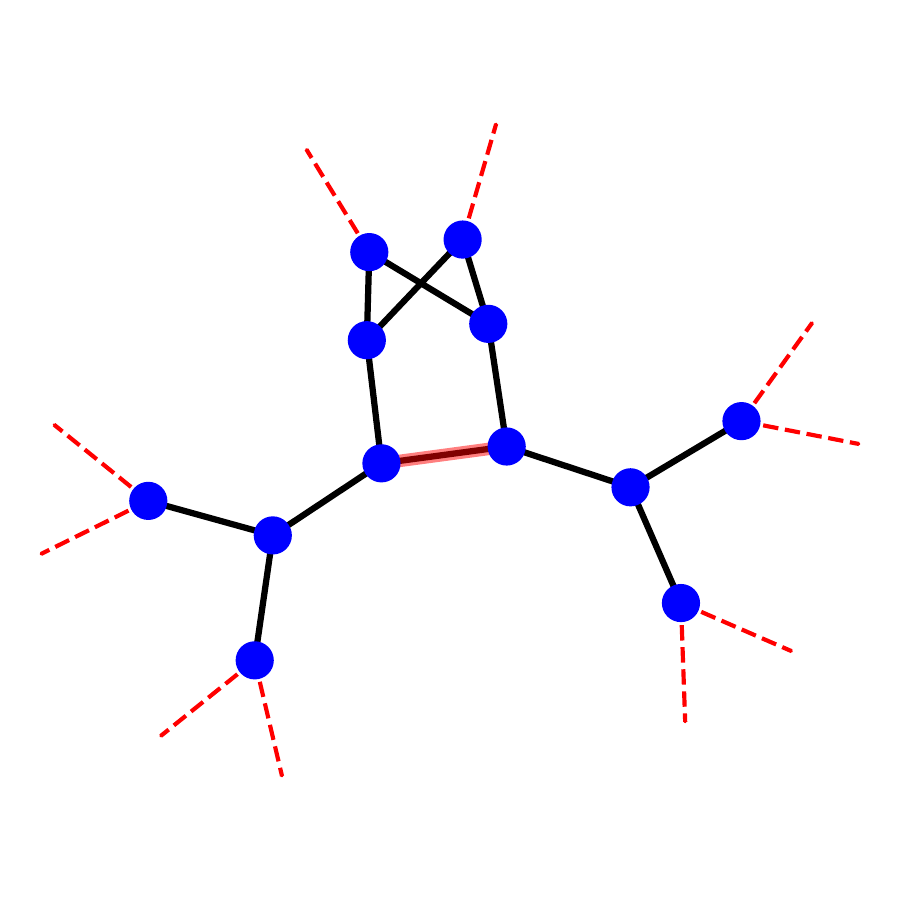}}
                & Index    & 2, 14 \\ \cline{2-3} 
                & $c_{14}$    & 0.9308 \\ \cline{2-3} 
                & $f_{14}$    & 0.7445 \\ \cline{2-3} 
                & Envs.    & 11 \\ \cline{2-3} 
                & $f_{14}'$    & 0.7468 \\ \cline{2-3} 
                & $\gamma_{1}$    & $-152^\circ$\\ \cline{2-3} 
                & $\beta_{1}$    & $-60^\circ$ \\ \cline{2-3} 
                & $\gamma_{2}$    & $51^\circ$ \\
\hline 
\end{tabular}
\end{table}
\begin{table}[]
\begin{tabular}{|c|c|c|}
\hline
\multirow{8}{*}{\includegraphics[height=90pt]{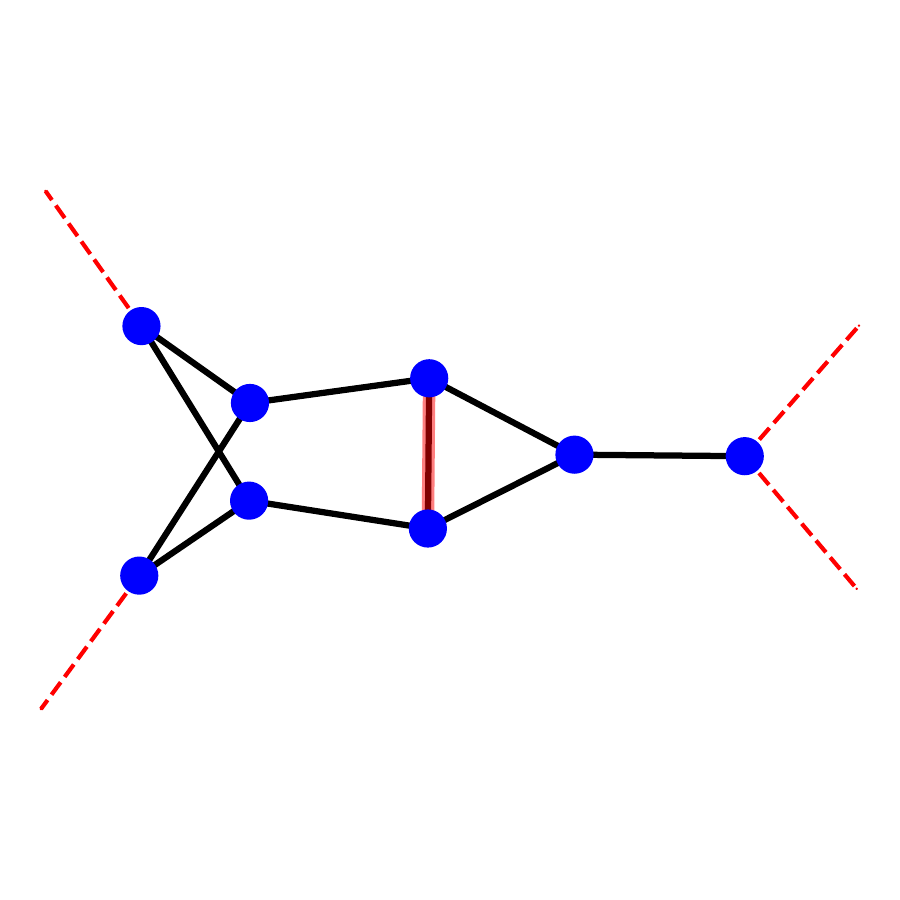}}
                & Index    & 2, 21 \\ \cline{2-3} 
                & $c_{21}$    & 0.9101 \\ \cline{2-3} 
                & $f_{21}$    & 0.6257 \\ \cline{2-3} 
                & Envs.    & 23 \\ \cline{2-3} 
                & $f_{21}'$    & 0.6590 \\ \cline{2-3} 
                & $\gamma_{1}$    & $147^\circ$\\ \cline{2-3} 
                & $\beta_{1}$    & $-14^\circ$ \\ \cline{2-3} 
                & $\gamma_{2}$    & $80^\circ$ \\
\hline 
\end{tabular}
\end{table}
\begin{table}[]
\begin{tabular}{|c|c|c|}
\hline
\multirow{8}{*}{\includegraphics[height=90pt]{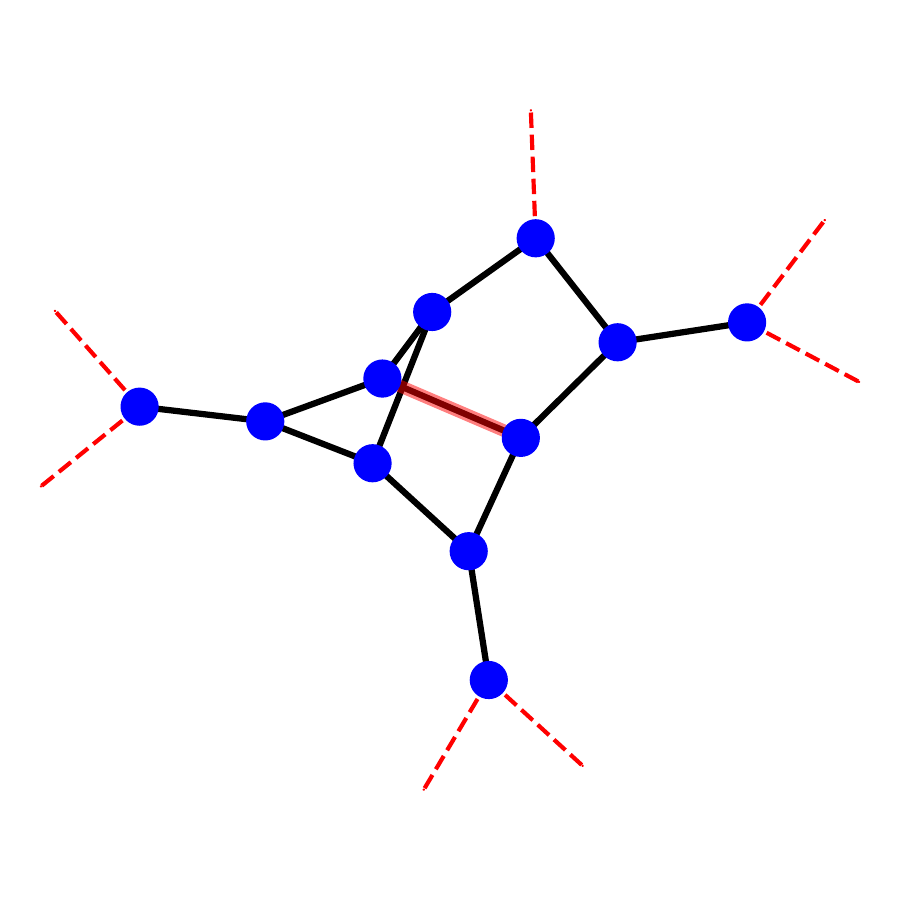}}
                & Index    & 2, 28 \\ \cline{2-3} 
                & $c_{28}$    & 0.9308 \\ \cline{2-3} 
                & $f_{28}$    & 0.7279 \\ \cline{2-3} 
                & Envs.    & 35 \\ \cline{2-3} 
                & $f_{28}'$    & 0.7358 \\ \cline{2-3} 
                & $\gamma_{1}$    & $26^\circ$\\ \cline{2-3} 
                & $\beta_{1}$    & $-28^\circ$ \\ \cline{2-3} 
                & $\gamma_{2}$    & $47^\circ$ \\
\hline 
\end{tabular}
\end{table}
\begin{table}[]
\begin{tabular}{|c|c|c|}
\hline
\multirow{8}{*}{\includegraphics[height=90pt]{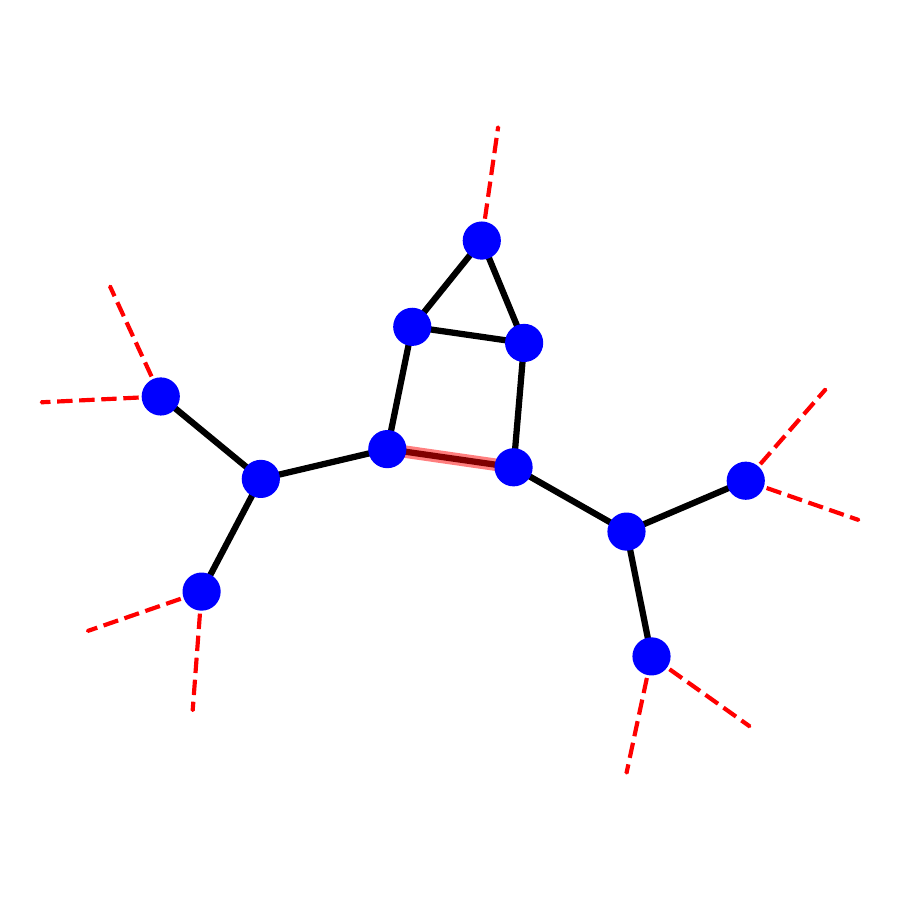}}
                & Index    & 2, 35 \\ \cline{2-3} 
                & $c_{35}$    & 0.9251 \\ \cline{2-3} 
                & $f_{35}$    & 0.7831 \\ \cline{2-3} 
                & Envs.    & 30 \\ \cline{2-3} 
                & $f_{35}'$    & 0.7852 \\ \cline{2-3} 
                & $\gamma_{1}$    & $152^\circ$\\ \cline{2-3} 
                & $\beta_{1}$    & $-33^\circ$ \\ \cline{2-3} 
                & $\gamma_{2}$    & $130^\circ$ \\
\hline 
\end{tabular}
\end{table}
\begin{table}[]
\begin{tabular}{|c|c|c|}
\hline
\multirow{8}{*}{\includegraphics[height=90pt]{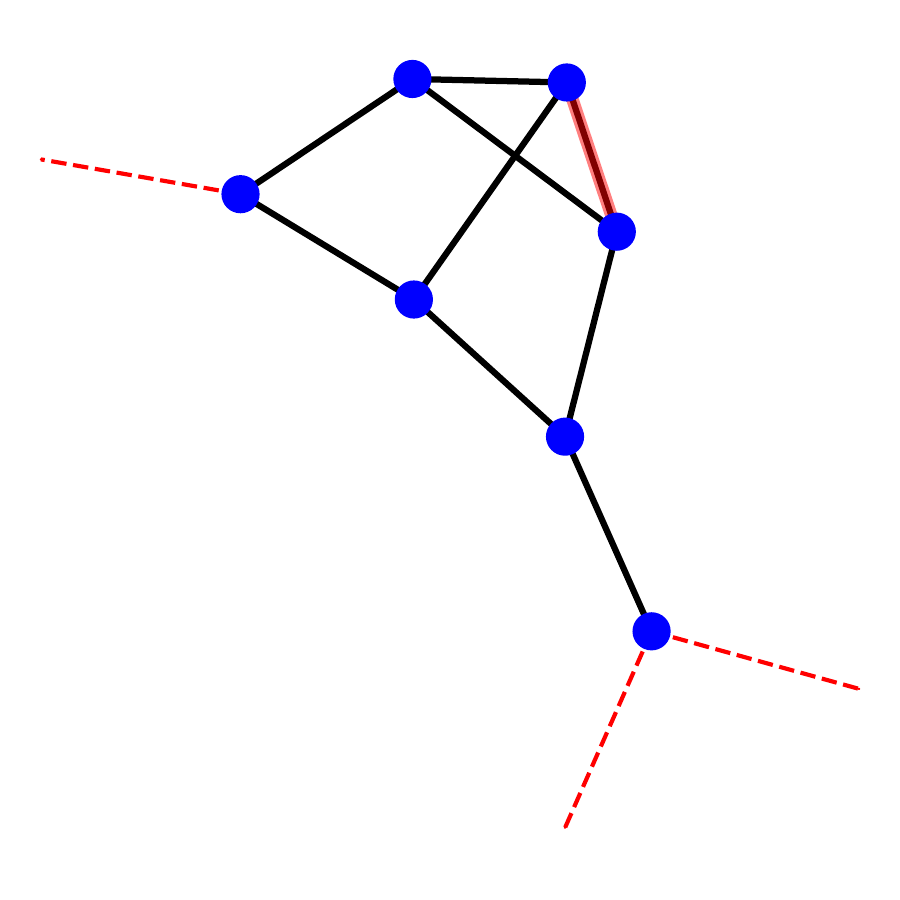}}
                & Index    & 2, 42 \\ \cline{2-3} 
                & $c_{42}$    & 0.9001 \\ \cline{2-3} 
                & $f_{42}$    & 0.6823 \\ \cline{2-3} 
                & Envs.    & 9 \\ \cline{2-3} 
                & $f_{42}'$    & 0.6885 \\ \cline{2-3} 
                & $\gamma_{1}$    & $35^\circ$\\ \cline{2-3} 
                & $\beta_{1}$    & $-28^\circ$ \\ \cline{2-3} 
                & $\gamma_{2}$    & $-126^\circ$ \\
\hline 
\end{tabular}
\end{table}
\begin{table}[]
\begin{tabular}{|c|c|c|}
\hline
\multirow{8}{*}{\includegraphics[height=90pt]{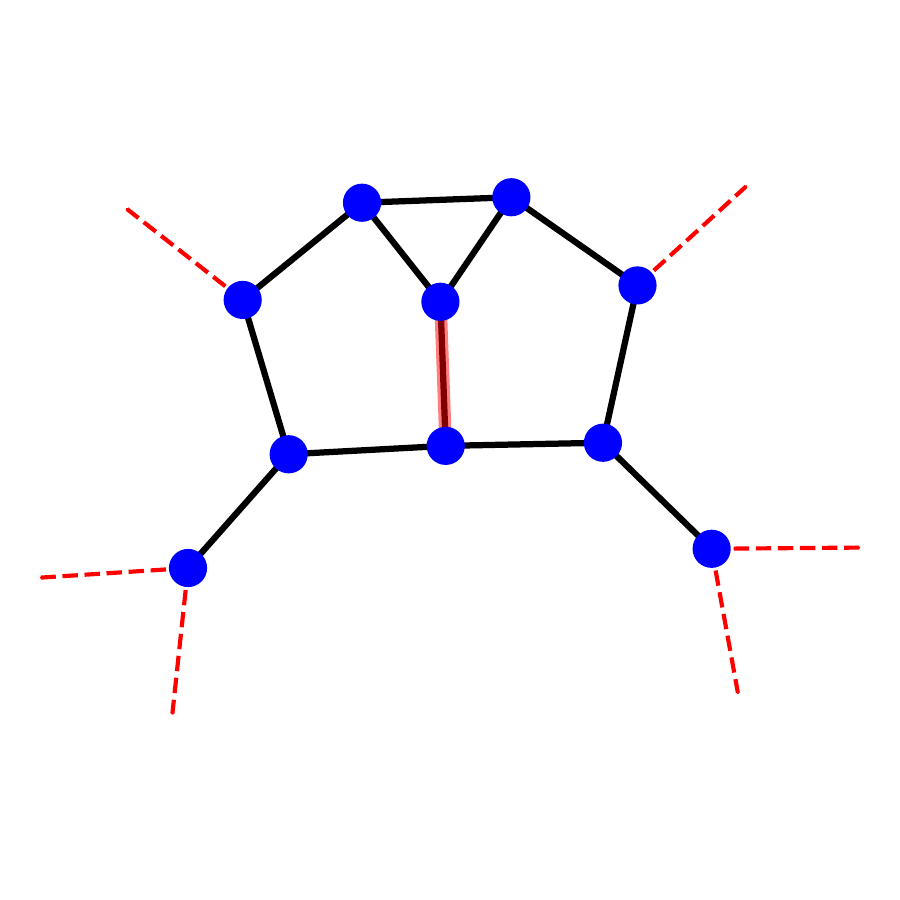}}
                & Index    & 2, 49 \\ \cline{2-3} 
                & $c_{49}$    & 0.8501 \\ \cline{2-3} 
                & $f_{49}$    & 0.7779 \\ \cline{2-3} 
                & Envs.    & 56 \\ \cline{2-3} 
                & $f_{49}'$    & 0.7954 \\ \cline{2-3} 
                & $\gamma_{1}$    & $146^\circ$\\ \cline{2-3} 
                & $\beta_{1}$    & $-27^\circ$ \\ \cline{2-3} 
                & $\gamma_{2}$    & $-61^\circ$ \\
\hline 
\end{tabular}
\end{table}
\begin{table}[]
\begin{tabular}{|c|c|c|}
\hline
\multirow{8}{*}{\includegraphics[height=90pt]{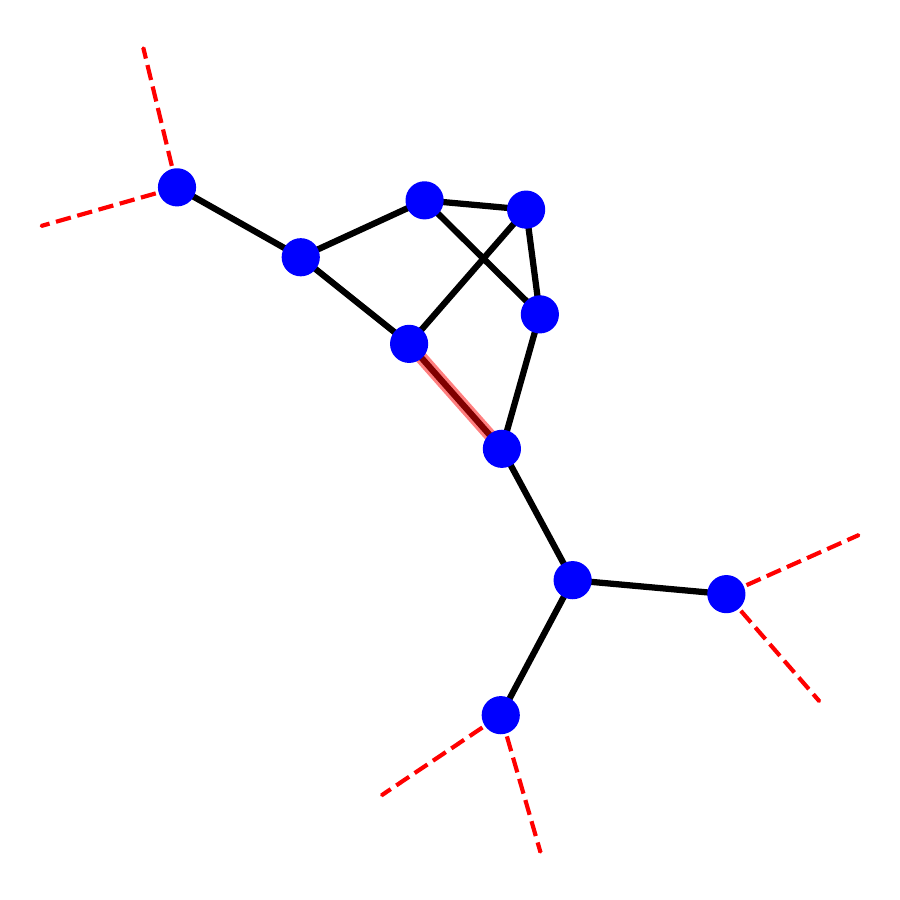}}
                & Index    & 2, 56 \\ \cline{2-3} 
                & $c_{56}$    & 0.9251 \\ \cline{2-3} 
                & $f_{56}$    & 0.7645 \\ \cline{2-3} 
                & Envs.    & 5 \\ \cline{2-3} 
                & $f_{56}'$    & 0.7679 \\ \cline{2-3} 
                & $\gamma_{1}$    & $26^\circ$\\ \cline{2-3} 
                & $\beta_{1}$    & $59^\circ$ \\ \cline{2-3} 
                & $\gamma_{2}$    & $47^\circ$ \\
\hline 
\end{tabular}
\end{table}
\begin{table}[]
\begin{tabular}{|c|c|c|}
\hline
\multirow{8}{*}{\includegraphics[height=90pt]{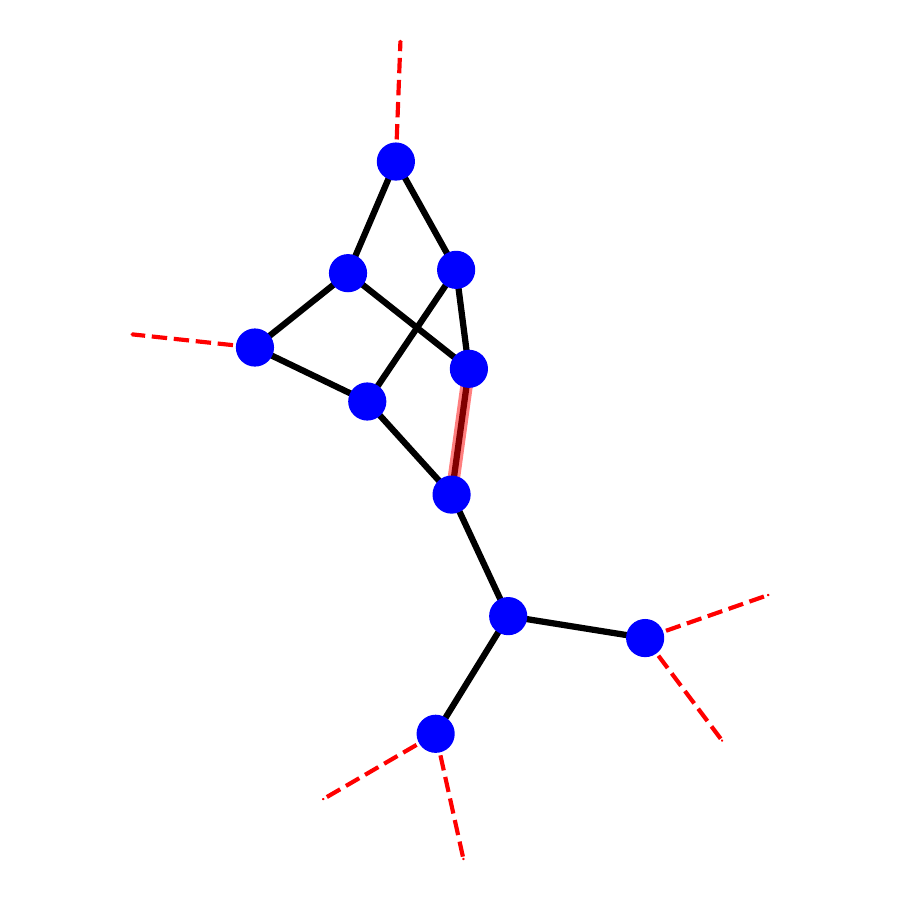}}
                & Index    & 2, 63 \\ \cline{2-3} 
                & $c_{63}$    & 0.9251 \\ \cline{2-3} 
                & $f_{63}$    & 0.7722 \\ \cline{2-3} 
                & Envs.    & 27 \\ \cline{2-3} 
                & $f_{63}'$    & 0.7758 \\ \cline{2-3} 
                & $\gamma_{1}$    & $153^\circ$\\ \cline{2-3} 
                & $\beta_{1}$    & $-32^\circ$ \\ \cline{2-3} 
                & $\gamma_{2}$    & $-48^\circ$ \\
\hline 
\end{tabular}
\end{table}
\begin{table}[]
\begin{tabular}{|c|c|c|}
\hline
\multirow{8}{*}{\includegraphics[height=90pt]{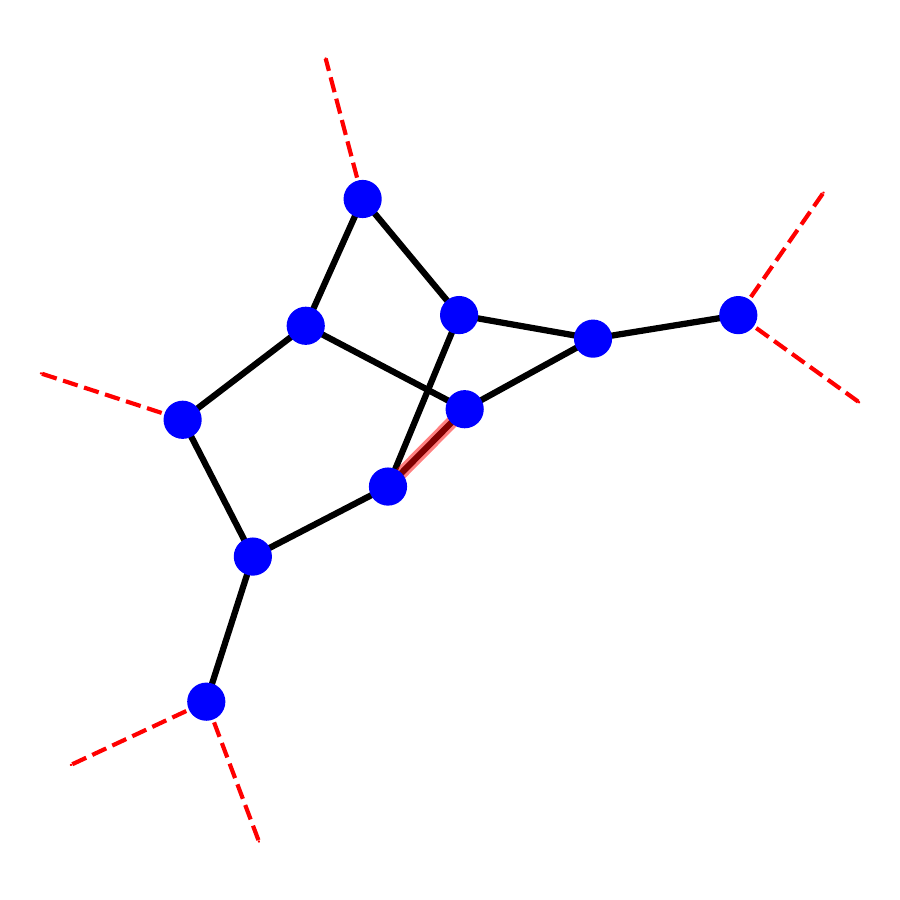}}
                & Index    & 2, 70 \\ \cline{2-3} 
                & $c_{70}$    & 0.9251 \\ \cline{2-3} 
                & $f_{70}$    & 0.7799 \\ \cline{2-3} 
                & Envs.    & 136 \\ \cline{2-3} 
                & $f_{70}'$    & 0.7811 \\ \cline{2-3} 
                & $\gamma_{1}$    & $-208^\circ$\\ \cline{2-3} 
                & $\beta_{1}$    & $58^\circ$ \\ \cline{2-3} 
                & $\gamma_{2}$    & $130^\circ$ \\
\hline 
\end{tabular}
\end{table}
\begin{table}[]
\begin{tabular}{|c|c|c|}
\hline
\multirow{8}{*}{\includegraphics[height=90pt]{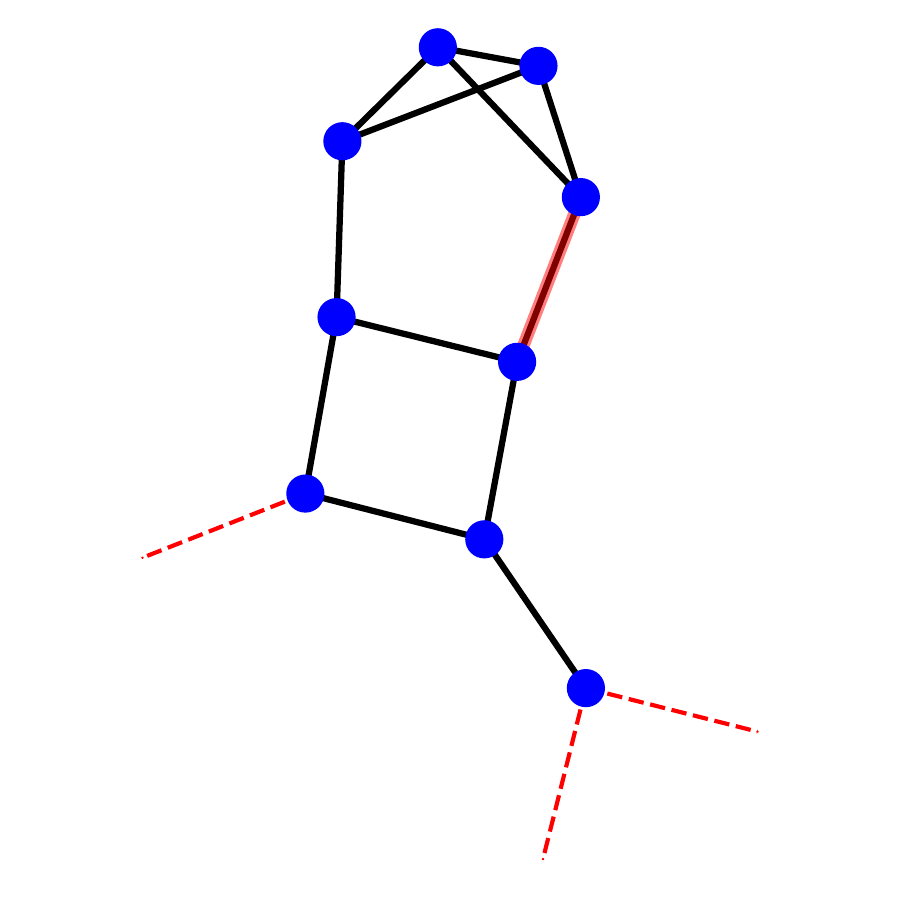}}
                & Index    & 2, 77 \\ \cline{2-3} 
                & $c_{77}$    & 0.8501 \\ \cline{2-3} 
                & $f_{77}$    & 0.7552 \\ \cline{2-3} 
                & Envs.    & 2 \\ \cline{2-3} 
                & $f_{77}'$    & 0.7674 \\ \cline{2-3} 
                & $\gamma_{1}$    & $208^\circ$\\ \cline{2-3} 
                & $\beta_{1}$    & $-64^\circ$ \\ \cline{2-3} 
                & $\gamma_{2}$    & $52^\circ$ \\
\hline 
\end{tabular}
\end{table}
\begin{table}[]
\begin{tabular}{|c|c|c|}
\hline
\multirow{8}{*}{\includegraphics[height=90pt]{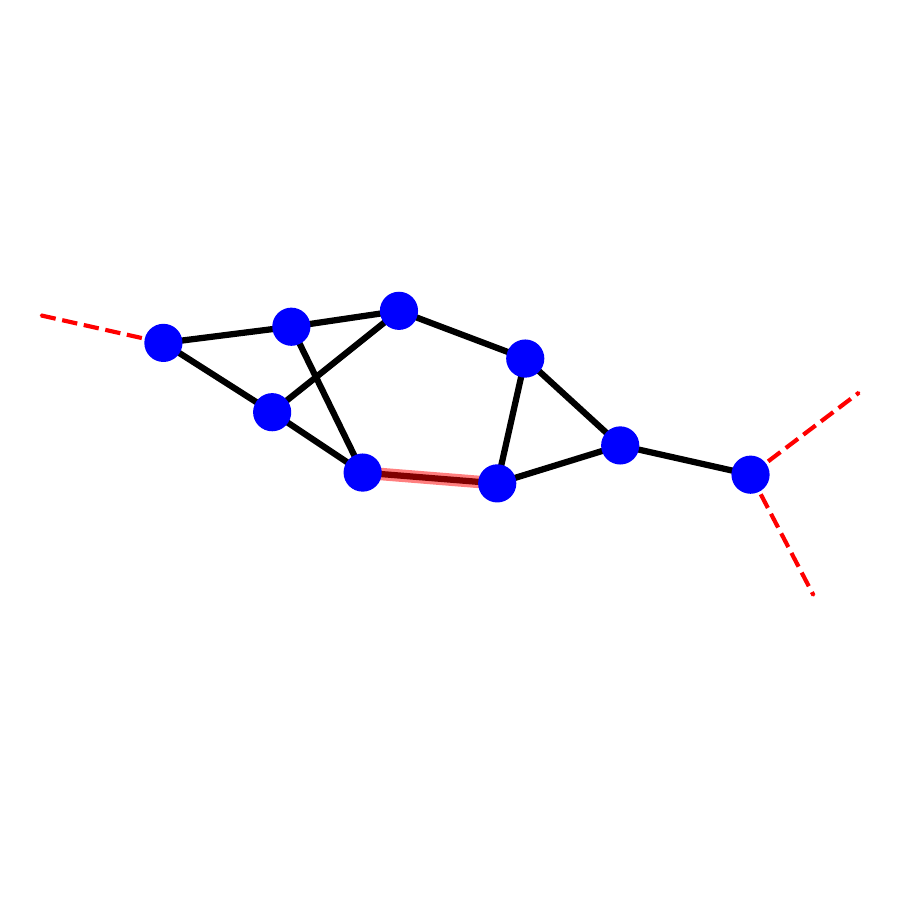}}
                & Index    & 2, 84 \\ \cline{2-3} 
                & $c_{84}$    & 0.9251 \\ \cline{2-3} 
                & $f_{84}$    & 0.7605 \\ \cline{2-3} 
                & Envs.    & 1 \\ \cline{2-3} 
                & $f_{84}'$    & 0.7726 \\ \cline{2-3} 
                & $\gamma_{1}$    & $-149^\circ$\\ \cline{2-3} 
                & $\beta_{1}$    & $-63^\circ$ \\ \cline{2-3} 
                & $\gamma_{2}$    & $-126^\circ$ \\
\hline 
\end{tabular}
\end{table}
\begin{table}[]
\begin{tabular}{|c|c|c|}
\hline
\multirow{8}{*}{\includegraphics[height=90pt]{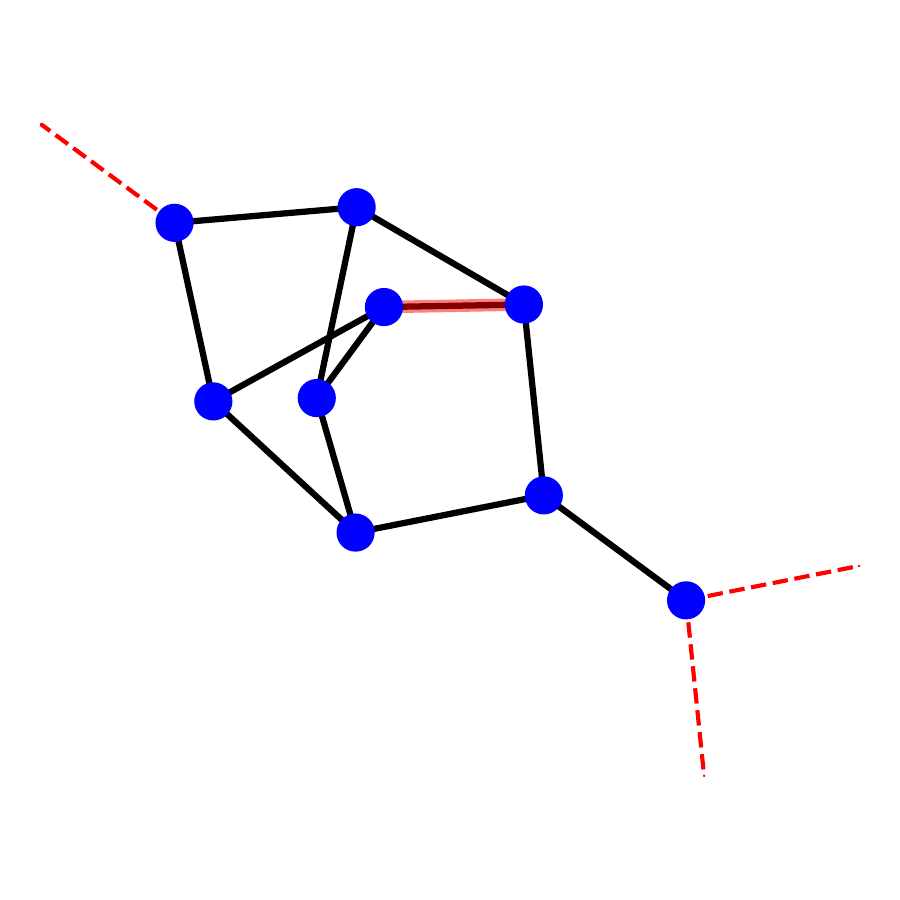}}
                & Index    & 2, 91 \\ \cline{2-3} 
                & $c_{91}$    & 0.8501 \\ \cline{2-3} 
                & $f_{91}$    & 0.7615 \\ \cline{2-3} 
                & Envs.    & 2 \\ \cline{2-3} 
                & $f_{91}'$    & 0.7647 \\ \cline{2-3} 
                & $\gamma_{1}$    & $154^\circ$\\ \cline{2-3} 
                & $\beta_{1}$    & $-31^\circ$ \\ \cline{2-3} 
                & $\gamma_{2}$    & $133^\circ$ \\
\hline 
\end{tabular}
\end{table}
\begin{table}[]
\begin{tabular}{|c|c|c|}
\hline
\multirow{8}{*}{\includegraphics[height=90pt]{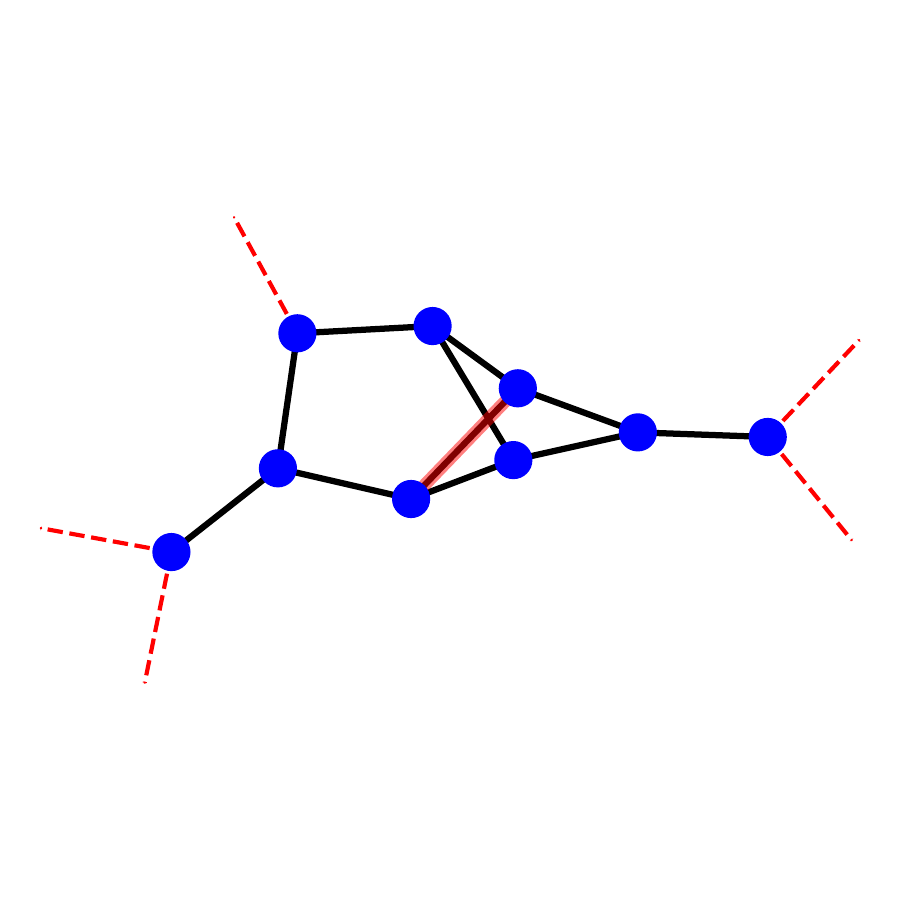}}
                & Index    & 2, 98 \\ \cline{2-3} 
                & $c_{98}$    & 0.9182 \\ \cline{2-3} 
                & $f_{98}$    & 0.8109 \\ \cline{2-3} 
                & Envs.    & 33 \\ \cline{2-3} 
                & $f_{98}'$    & 0.8256 \\ \cline{2-3} 
                & $\gamma_{1}$    & $-29^\circ$\\ \cline{2-3} 
                & $\beta_{1}$    & $34^\circ$ \\ \cline{2-3} 
                & $\gamma_{2}$    & $132^\circ$ \\
\hline 
\end{tabular}
\end{table}
\begin{table}[]
\begin{tabular}{|c|c|c|}
\hline
\multirow{8}{*}{\includegraphics[height=90pt]{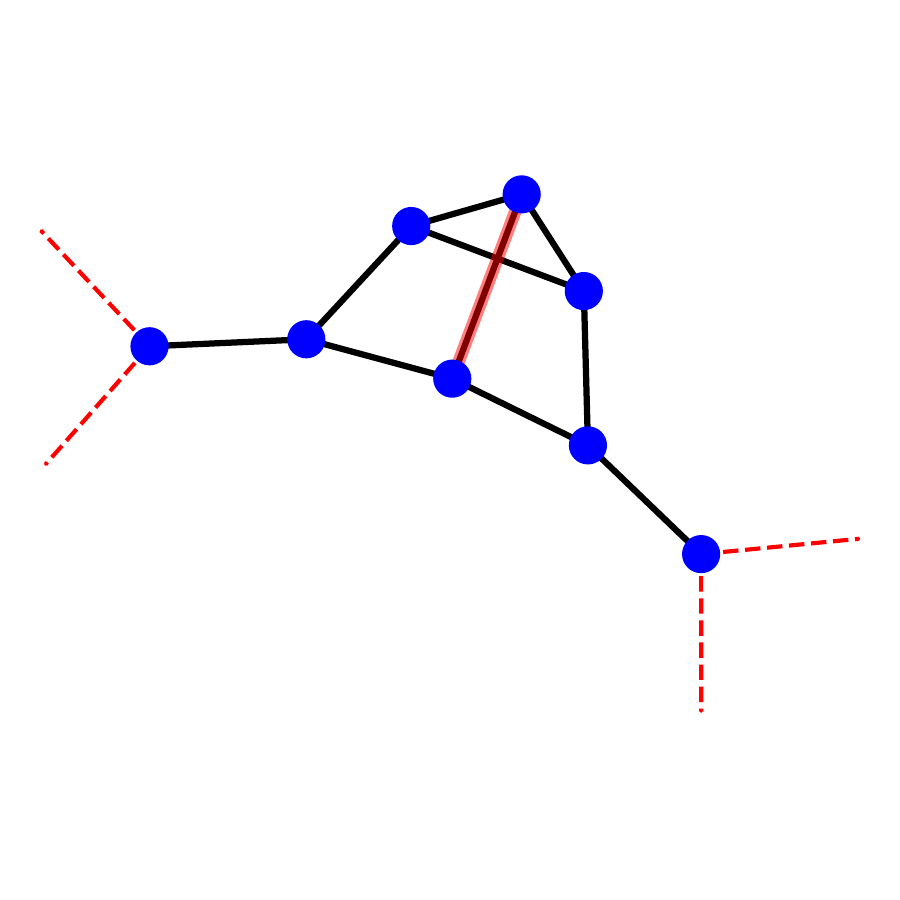}}
                & Index    & 2, 105 \\ \cline{2-3} 
                & $c_{105}$    & 0.9101 \\ \cline{2-3} 
                & $f_{105}$    & 0.8522 \\ \cline{2-3} 
                & Envs.    & 18 \\ \cline{2-3} 
                & $f_{105}'$    & 0.8608 \\ \cline{2-3} 
                & $\gamma_{1}$    & $-31^\circ$\\ \cline{2-3} 
                & $\beta_{1}$    & $32^\circ$ \\ \cline{2-3} 
                & $\gamma_{2}$    & $-53^\circ$ \\
\hline 
\end{tabular}
\end{table}
\begin{table}[]
\begin{tabular}{|c|c|c|}
\hline
\multirow{8}{*}{\includegraphics[height=90pt]{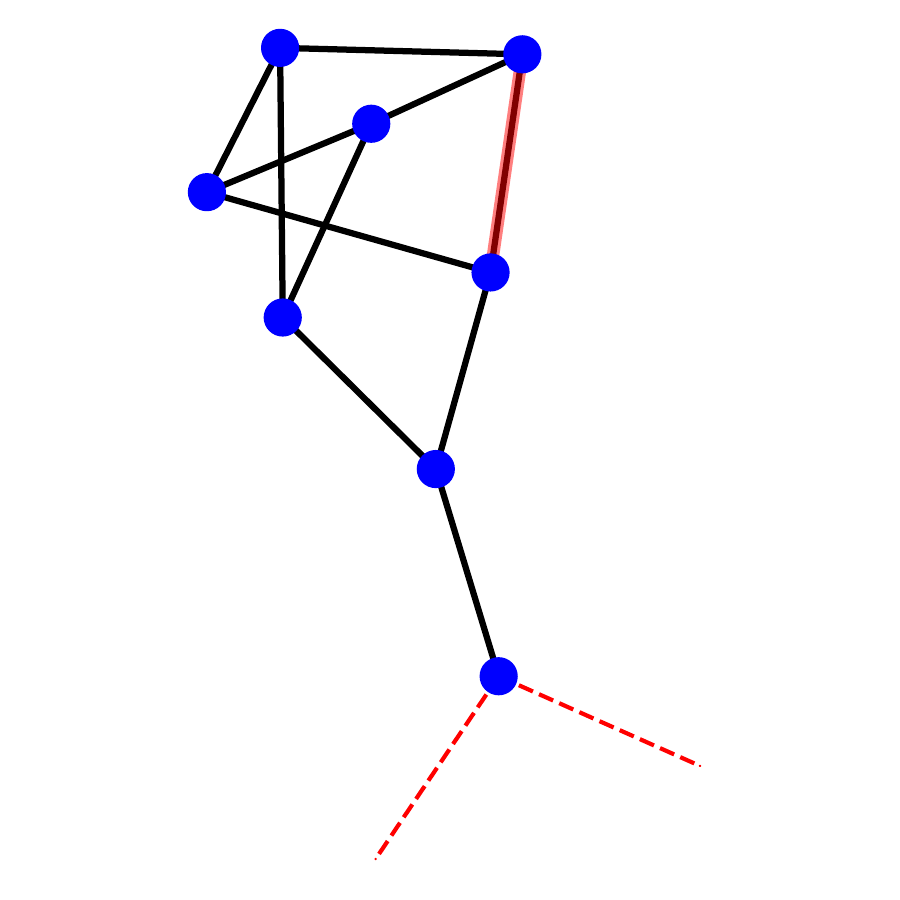}}
                & Index    & 2, 112 \\ \cline{2-3} 
                & $c_{112}$    & 0.9182 \\ \cline{2-3} 
                & $f_{112}$    & 0.7930 \\ \cline{2-3} 
                & Envs.    & 1 \\ \cline{2-3} 
                & $f_{112}'$    & 0.8051 \\ \cline{2-3} 
                & $\gamma_{1}$    & $152^\circ$\\ \cline{2-3} 
                & $\beta_{1}$    & $-33^\circ$ \\ \cline{2-3} 
                & $\gamma_{2}$    & $-46^\circ$ \\
\hline 
\end{tabular}
\end{table}
\begin{table}[]
\begin{tabular}{|c|c|c|}
\hline
\multirow{8}{*}{\includegraphics[height=90pt]{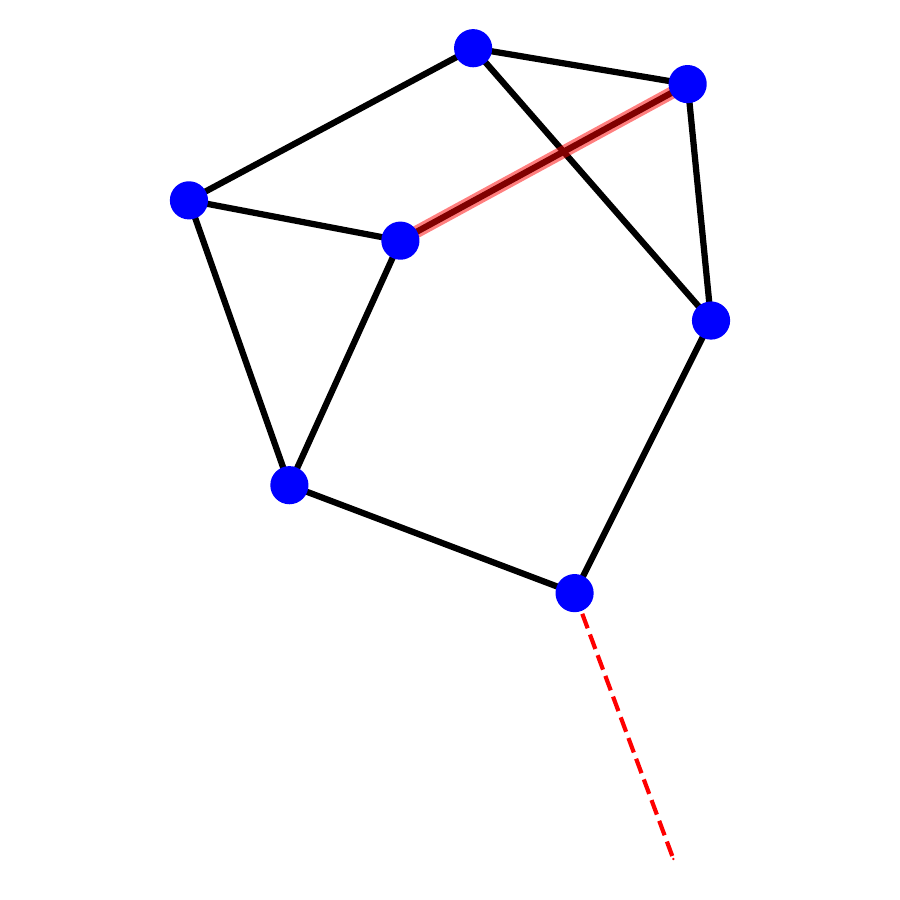}}
                & Index    & 2, 119 \\ \cline{2-3} 
                & $c_{119}$    & 0.8201 \\ \cline{2-3} 
                & $f_{119}$    & 0.8480 \\ \cline{2-3} 
                & Envs.    & 1 \\ \cline{2-3} 
                & $f_{119}'$    & 0.8799 \\ \cline{2-3} 
                & $\gamma_{1}$    & $35^\circ$\\ \cline{2-3} 
                & $\beta_{1}$    & $-27^\circ$ \\ \cline{2-3} 
                & $\gamma_{2}$    & $62^\circ$ \\
\hline 
\end{tabular}
\end{table}

%% file: main.bbl
\begin{thebibliography}{26}%
\makeatletter
\providecommand \@ifxundefined [1]{%
 \@ifx{#1\undefined}
}%
\providecommand \@ifnum [1]{%
 \ifnum #1\expandafter \@firstoftwo
 \else \expandafter \@secondoftwo
 \fi
}%
\providecommand \@ifx [1]{%
 \ifx #1\expandafter \@firstoftwo
 \else \expandafter \@secondoftwo
 \fi
}%
\providecommand \natexlab [1]{#1}%
\providecommand \enquote  [1]{``#1''}%
\providecommand \bibnamefont  [1]{#1}%
\providecommand \bibfnamefont [1]{#1}%
\providecommand \citenamefont [1]{#1}%
\providecommand \href@noop [0]{\@secondoftwo}%
\providecommand \href [0]{\begingroup \@sanitize@url \@href}%
\providecommand \@href[1]{\@@startlink{#1}\@@href}%
\providecommand \@@href[1]{\endgroup#1\@@endlink}%
\providecommand \@sanitize@url [0]{\catcode `\\12\catcode `\$12\catcode
  `\&12\catcode `\#12\catcode `\^12\catcode `\_12\catcode `\%12\relax}%
\providecommand \@@startlink[1]{}%
\providecommand \@@endlink[0]{}%
\providecommand \url  [0]{\begingroup\@sanitize@url \@url }%
\providecommand \@url [1]{\endgroup\@href {#1}{\urlprefix }}%
\providecommand \urlprefix  [0]{URL }%
\providecommand \Eprint [0]{\href }%
\providecommand \doibase [0]{http://dx.doi.org/}%
\providecommand \selectlanguage [0]{\@gobble}%
\providecommand \bibinfo  [0]{\@secondoftwo}%
\providecommand \bibfield  [0]{\@secondoftwo}%
\providecommand \translation [1]{[#1]}%
\providecommand \BibitemOpen [0]{}%
\providecommand \bibitemStop [0]{}%
\providecommand \bibitemNoStop [0]{.\EOS\space}%
\providecommand \EOS [0]{\spacefactor3000\relax}%
\providecommand \BibitemShut  [1]{\csname bibitem#1\endcsname}%
\let\auto@bib@innerbib\@empty
\bibitem [{\citenamefont {Preskill}(2018)}]{Preskill2018}%
  \BibitemOpen
  \bibfield  {author} {\bibinfo {author} {\bibfnamefont {J.}~\bibnamefont
  {Preskill}},\ }\href {\doibase 10.22331/q-2018-08-06-79} {\bibfield
  {journal} {\bibinfo  {journal} {{Quantum}}\ }\textbf {\bibinfo {volume}
  {2}},\ \bibinfo {pages} {79} (\bibinfo {year} {2018})}\BibitemShut {NoStop}%
\bibitem [{\citenamefont {Farhi}\ \emph
  {et~al.}(2014{\natexlab{a}})\citenamefont {Farhi}, \citenamefont
  {Goldstone},\ and\ \citenamefont {Gutmann}}]{farhi2014}%
  \BibitemOpen
  \bibfield  {author} {\bibinfo {author} {\bibfnamefont {E.}~\bibnamefont
  {Farhi}}, \bibinfo {author} {\bibfnamefont {J.}~\bibnamefont {Goldstone}}, \
  and\ \bibinfo {author} {\bibfnamefont {S.}~\bibnamefont {Gutmann}},\
  }\href@noop {} {\  (\bibinfo {year} {2014}{\natexlab{a}})},\ \Eprint
  {http://arxiv.org/abs/1411.4028} {arXiv:1411.4028 [quant-ph]} \BibitemShut
  {NoStop}%
\bibitem [{\citenamefont {Crooks}(2018)}]{crooks2018}%
  \BibitemOpen
  \bibfield  {author} {\bibinfo {author} {\bibfnamefont {G.~E.}\ \bibnamefont
  {Crooks}},\ }\href@noop {} {\  (\bibinfo {year} {2018})},\ \Eprint
  {http://arxiv.org/abs/1811.08419} {arXiv:1811.08419 [quant-ph]} \BibitemShut
  {NoStop}%
\bibitem [{\citenamefont {Hastings}(2019)}]{hastings2019classical}%
  \BibitemOpen
  \bibfield  {author} {\bibinfo {author} {\bibfnamefont {M.~B.}\ \bibnamefont
  {Hastings}},\ }\href@noop {} {\  (\bibinfo {year} {2019})},\ \Eprint
  {http://arxiv.org/abs/1905.07047} {arXiv:1905.07047 [quant-ph]} \BibitemShut
  {NoStop}%
\bibitem [{\citenamefont {Farhi}\ \emph
  {et~al.}(2014{\natexlab{b}})\citenamefont {Farhi}, \citenamefont
  {Goldstone},\ and\ \citenamefont {Gutmann}}]{farhi201e3lin0}%
  \BibitemOpen
  \bibfield  {author} {\bibinfo {author} {\bibfnamefont {E.}~\bibnamefont
  {Farhi}}, \bibinfo {author} {\bibfnamefont {J.}~\bibnamefont {Goldstone}}, \
  and\ \bibinfo {author} {\bibfnamefont {S.}~\bibnamefont {Gutmann}},\
  }\href@noop {} {\  (\bibinfo {year} {2014}{\natexlab{b}})},\ \Eprint
  {http://arxiv.org/abs/1412.6062v1} {arXiv:1412.6062v1 [quant-ph]}
  \BibitemShut {NoStop}%
\bibitem [{\citenamefont {Barak}\ \emph {et~al.}(2015)\citenamefont {Barak},
  \citenamefont {Moitra}, \citenamefont {O'Donnell}, \citenamefont
  {Raghavendra}, \citenamefont {Regev}, \citenamefont {Steurer}, \citenamefont
  {Trevisan}, \citenamefont {Vijayaraghavan}, \citenamefont {Witmer},\ and\
  \citenamefont {Wright}}]{barak2015}%
  \BibitemOpen
  \bibfield  {author} {\bibinfo {author} {\bibfnamefont {B.}~\bibnamefont
  {Barak}}, \bibinfo {author} {\bibfnamefont {A.}~\bibnamefont {Moitra}},
  \bibinfo {author} {\bibfnamefont {R.}~\bibnamefont {O'Donnell}}, \bibinfo
  {author} {\bibfnamefont {P.}~\bibnamefont {Raghavendra}}, \bibinfo {author}
  {\bibfnamefont {O.}~\bibnamefont {Regev}}, \bibinfo {author} {\bibfnamefont
  {D.}~\bibnamefont {Steurer}}, \bibinfo {author} {\bibfnamefont
  {L.}~\bibnamefont {Trevisan}}, \bibinfo {author} {\bibfnamefont
  {A.}~\bibnamefont {Vijayaraghavan}}, \bibinfo {author} {\bibfnamefont
  {D.}~\bibnamefont {Witmer}}, \ and\ \bibinfo {author} {\bibfnamefont
  {J.}~\bibnamefont {Wright}},\ }\href@noop {} {\  (\bibinfo {year} {2015})},\
  \Eprint {http://arxiv.org/abs/1505.03424} {arXiv:1505.03424 [cs.CC]}
  \BibitemShut {NoStop}%
\bibitem [{\citenamefont {Farhi}\ \emph {et~al.}(2015)\citenamefont {Farhi},
  \citenamefont {Goldstone},\ and\ \citenamefont {Gutmann}}]{farhi201e3lin}%
  \BibitemOpen
  \bibfield  {author} {\bibinfo {author} {\bibfnamefont {E.}~\bibnamefont
  {Farhi}}, \bibinfo {author} {\bibfnamefont {J.}~\bibnamefont {Goldstone}}, \
  and\ \bibinfo {author} {\bibfnamefont {S.}~\bibnamefont {Gutmann}},\
  }\href@noop {} {\  (\bibinfo {year} {2015})},\ \Eprint
  {http://arxiv.org/abs/1412.6062v2} {arXiv:1412.6062v2 [quant-ph]}
  \BibitemShut {NoStop}%
\bibitem [{\citenamefont {Garey}\ \emph {et~al.}(1976)\citenamefont {Garey},
  \citenamefont {Johnson},\ and\ \citenamefont {Stockmeyer}}]{garey1976}%
  \BibitemOpen
  \bibfield  {author} {\bibinfo {author} {\bibfnamefont {M.}~\bibnamefont
  {Garey}}, \bibinfo {author} {\bibfnamefont {D.}~\bibnamefont {Johnson}}, \
  and\ \bibinfo {author} {\bibfnamefont {L.}~\bibnamefont {Stockmeyer}},\
  }\href {\doibase https://doi.org/10.1016/0304-3975(76)90059-1} {\bibfield
  {journal} {\bibinfo  {journal} {Theoretical Computer Science}\ }\textbf
  {\bibinfo {volume} {1}},\ \bibinfo {pages} {237 } (\bibinfo {year}
  {1976})}\BibitemShut {NoStop}%
\bibitem [{\citenamefont {Crosson}\ \emph {et~al.}(2014)\citenamefont
  {Crosson}, \citenamefont {Farhi}, \citenamefont {Lin}, \citenamefont {Lin},\
  and\ \citenamefont {Shor}}]{crosson2014}%
  \BibitemOpen
  \bibfield  {author} {\bibinfo {author} {\bibfnamefont {E.}~\bibnamefont
  {Crosson}}, \bibinfo {author} {\bibfnamefont {E.}~\bibnamefont {Farhi}},
  \bibinfo {author} {\bibfnamefont {C.~Y.-Y.}\ \bibnamefont {Lin}}, \bibinfo
  {author} {\bibfnamefont {H.-H.}\ \bibnamefont {Lin}}, \ and\ \bibinfo
  {author} {\bibfnamefont {P.}~\bibnamefont {Shor}},\ }\href@noop {} {\
  (\bibinfo {year} {2014})},\ \Eprint {http://arxiv.org/abs/1401.7320}
  {arXiv:1401.7320 [quant-ph]} \BibitemShut {NoStop}%
\bibitem [{\citenamefont {Wang}\ \emph {et~al.}(2018)\citenamefont {Wang},
  \citenamefont {Hadfield}, \citenamefont {Jiang},\ and\ \citenamefont
  {Rieffel}}]{Wang2018}%
  \BibitemOpen
  \bibfield  {author} {\bibinfo {author} {\bibfnamefont {Z.}~\bibnamefont
  {Wang}}, \bibinfo {author} {\bibfnamefont {S.}~\bibnamefont {Hadfield}},
  \bibinfo {author} {\bibfnamefont {Z.}~\bibnamefont {Jiang}}, \ and\ \bibinfo
  {author} {\bibfnamefont {E.~G.}\ \bibnamefont {Rieffel}},\ }\href {\doibase
  10.1103/PhysRevA.97.022304} {\bibfield  {journal} {\bibinfo  {journal} {Phys.
  Rev. A}\ }\textbf {\bibinfo {volume} {97}},\ \bibinfo {pages} {022304}
  (\bibinfo {year} {2018})}\BibitemShut {NoStop}%
\bibitem [{\citenamefont {Willsch}\ \emph {et~al.}(2020)\citenamefont
  {Willsch}, \citenamefont {Willsch}, \citenamefont {Jin}, \citenamefont
  {De~Raedt},\ and\ \citenamefont {Michielsen}}]{Willsch2020}%
  \BibitemOpen
  \bibfield  {author} {\bibinfo {author} {\bibfnamefont {M.}~\bibnamefont
  {Willsch}}, \bibinfo {author} {\bibfnamefont {D.}~\bibnamefont {Willsch}},
  \bibinfo {author} {\bibfnamefont {F.}~\bibnamefont {Jin}}, \bibinfo {author}
  {\bibfnamefont {H.}~\bibnamefont {De~Raedt}}, \ and\ \bibinfo {author}
  {\bibfnamefont {K.}~\bibnamefont {Michielsen}},\ }\href {\doibase
  10.1007/s11128-020-02692-8} {\bibfield  {journal} {\bibinfo  {journal}
  {Quantum Information Processing}\ }\textbf {\bibinfo {volume} {19}},\
  \bibinfo {pages} {197} (\bibinfo {year} {2020})}\BibitemShut {NoStop}%
\bibitem [{\citenamefont {Larkin}\ \emph {et~al.}(2020)\citenamefont {Larkin},
  \citenamefont {Jonsson}, \citenamefont {Justice},\ and\ \citenamefont
  {Guerreschi}}]{larkin2020}%
  \BibitemOpen
  \bibfield  {author} {\bibinfo {author} {\bibfnamefont {J.}~\bibnamefont
  {Larkin}}, \bibinfo {author} {\bibfnamefont {M.}~\bibnamefont {Jonsson}},
  \bibinfo {author} {\bibfnamefont {D.}~\bibnamefont {Justice}}, \ and\
  \bibinfo {author} {\bibfnamefont {G.~G.}\ \bibnamefont {Guerreschi}},\
  }\href@noop {} {\  (\bibinfo {year} {2020})},\ \Eprint
  {http://arxiv.org/abs/2006.04831} {arXiv:2006.04831 [quant-ph]} \BibitemShut
  {NoStop}%
\bibitem [{\citenamefont {Zhou}\ \emph {et~al.}(2020)\citenamefont {Zhou},
  \citenamefont {Wang}, \citenamefont {Choi}, \citenamefont {Pichler},\ and\
  \citenamefont {Lukin}}]{Zhou2020}%
  \BibitemOpen
  \bibfield  {author} {\bibinfo {author} {\bibfnamefont {L.}~\bibnamefont
  {Zhou}}, \bibinfo {author} {\bibfnamefont {S.-T.}\ \bibnamefont {Wang}},
  \bibinfo {author} {\bibfnamefont {S.}~\bibnamefont {Choi}}, \bibinfo {author}
  {\bibfnamefont {H.}~\bibnamefont {Pichler}}, \ and\ \bibinfo {author}
  {\bibfnamefont {M.~D.}\ \bibnamefont {Lukin}},\ }\href {\doibase
  10.1103/PhysRevX.10.021067} {\bibfield  {journal} {\bibinfo  {journal} {Phys.
  Rev. X}\ }\textbf {\bibinfo {volume} {10}},\ \bibinfo {pages} {021067}
  (\bibinfo {year} {2020})}\BibitemShut {NoStop}%
\bibitem [{\citenamefont {{Google AI Quantum and
  Collaborators}}(2020)}]{arute2020quantum}%
  \BibitemOpen
  \bibfield  {author} {\bibinfo {author} {\bibnamefont {{Google AI Quantum and
  Collaborators}}},\ }\href@noop {} {\  (\bibinfo {year} {2020})},\ \Eprint
  {http://arxiv.org/abs/2004.04197} {arXiv:2004.04197 [quant-ph]} \BibitemShut
  {NoStop}%
\bibitem [{\citenamefont {Paton}(1969)}]{Paton1969}%
  \BibitemOpen
  \bibfield  {author} {\bibinfo {author} {\bibfnamefont {K.}~\bibnamefont
  {Paton}},\ }\href@noop {} {\bibfield  {journal} {\bibinfo  {journal} {Commun.
  ACM}\ }\textbf {\bibinfo {volume} {12}},\ \bibinfo {pages} {514} (\bibinfo
  {year} {1969})}\BibitemShut {NoStop}%
\bibitem [{\citenamefont {Zaslavsky}(2018)}]{ZASLAVSKY2018}%
  \BibitemOpen
  \bibfield  {author} {\bibinfo {author} {\bibfnamefont {T.}~\bibnamefont
  {Zaslavsky}},\ }\href {\doibase https://doi.org/10.1016/j.akcej.2018.01.011}
  {\bibfield  {journal} {\bibinfo  {journal} {AKCE International Journal of
  Graphs and Combinatorics}\ }\textbf {\bibinfo {volume} {15}},\ \bibinfo
  {pages} {31 } (\bibinfo {year} {2018})},\ \bibinfo {note} {international
  Conference on Current trends in Graph Theory and Computation}\BibitemShut
  {NoStop}%
\bibitem [{\citenamefont {Farhi}\ \emph {et~al.}(2020)\citenamefont {Farhi},
  \citenamefont {Gamarnik},\ and\ \citenamefont {Gutmann}}]{farhi2020}%
  \BibitemOpen
  \bibfield  {author} {\bibinfo {author} {\bibfnamefont {E.}~\bibnamefont
  {Farhi}}, \bibinfo {author} {\bibfnamefont {D.}~\bibnamefont {Gamarnik}}, \
  and\ \bibinfo {author} {\bibfnamefont {S.}~\bibnamefont {Gutmann}},\
  }\href@noop {} {\  (\bibinfo {year} {2020})},\ \Eprint
  {http://arxiv.org/abs/2005.08747} {arXiv:2005.08747 [quant-ph]} \BibitemShut
  {NoStop}%
\bibitem [{\citenamefont {Brandao}\ \emph {et~al.}(2018)\citenamefont
  {Brandao}, \citenamefont {Broughton}, \citenamefont {Farhi}, \citenamefont
  {Gutmann},\ and\ \citenamefont {Neven}}]{brandao2018}%
  \BibitemOpen
  \bibfield  {author} {\bibinfo {author} {\bibfnamefont {F.~G. S.~L.}\
  \bibnamefont {Brandao}}, \bibinfo {author} {\bibfnamefont {M.}~\bibnamefont
  {Broughton}}, \bibinfo {author} {\bibfnamefont {E.}~\bibnamefont {Farhi}},
  \bibinfo {author} {\bibfnamefont {S.}~\bibnamefont {Gutmann}}, \ and\
  \bibinfo {author} {\bibfnamefont {H.}~\bibnamefont {Neven}},\ }\href@noop {}
  {\  (\bibinfo {year} {2018})},\ \Eprint {http://arxiv.org/abs/1812.04170}
  {arXiv:1812.04170 [quant-ph]} \BibitemShut {NoStop}%
\bibitem [{\citenamefont {Sahni}\ and\ \citenamefont
  {Gonzalez}(1976)}]{Sahni1976}%
  \BibitemOpen
  \bibfield  {author} {\bibinfo {author} {\bibfnamefont {S.}~\bibnamefont
  {Sahni}}\ and\ \bibinfo {author} {\bibfnamefont {T.}~\bibnamefont
  {Gonzalez}},\ }\href {\doibase 10.1145/321958.321975} {\bibfield  {journal}
  {\bibinfo  {journal} {J. ACM}\ }\textbf {\bibinfo {volume} {23}},\ \bibinfo
  {pages} {555–565} (\bibinfo {year} {1976})}\BibitemShut {NoStop}%
\bibitem [{\citenamefont {H\r{a}stad}(2001)}]{Hrastad2001}%
  \BibitemOpen
  \bibfield  {author} {\bibinfo {author} {\bibfnamefont {J.}~\bibnamefont
  {H\r{a}stad}},\ }\href {\doibase 10.1145/502090.502098} {\bibfield  {journal}
  {\bibinfo  {journal} {J. ACM}\ }\textbf {\bibinfo {volume} {48}},\ \bibinfo
  {pages} {798–859} (\bibinfo {year} {2001})}\BibitemShut {NoStop}%
\bibitem [{\citenamefont {Goemans}\ and\ \citenamefont
  {Williamson}(1995)}]{goemans1995}%
  \BibitemOpen
  \bibfield  {author} {\bibinfo {author} {\bibfnamefont {M.~X.}\ \bibnamefont
  {Goemans}}\ and\ \bibinfo {author} {\bibfnamefont {D.~P.}\ \bibnamefont
  {Williamson}},\ }\href {\doibase 10.1145/227683.227684} {\bibfield  {journal}
  {\bibinfo  {journal} {J. ACM}\ }\textbf {\bibinfo {volume} {42}},\ \bibinfo
  {pages} {1115–1145} (\bibinfo {year} {1995})}\BibitemShut {NoStop}%
\bibitem [{\citenamefont {Deza}\ and\ \citenamefont
  {Laurent}(1994)}]{Deza1994}%
  \BibitemOpen
  \bibfield  {author} {\bibinfo {author} {\bibfnamefont {M.}~\bibnamefont
  {Deza}}\ and\ \bibinfo {author} {\bibfnamefont {M.}~\bibnamefont {Laurent}},\
  }\href {\doibase https://doi.org/10.1016/0377-0427(94)90021-3} {\bibfield
  {journal} {\bibinfo  {journal} {Journal of Computational and Applied
  Mathematics}\ }\textbf {\bibinfo {volume} {55}},\ \bibinfo {pages} {217 }
  (\bibinfo {year} {1994})}\BibitemShut {NoStop}%
\bibitem [{\citenamefont {Halperin}\ \emph {et~al.}(2004)\citenamefont
  {Halperin}, \citenamefont {Livnat},\ and\ \citenamefont
  {Zwick}}]{Halperin2004}%
  \BibitemOpen
  \bibfield  {author} {\bibinfo {author} {\bibfnamefont {E.}~\bibnamefont
  {Halperin}}, \bibinfo {author} {\bibfnamefont {D.}~\bibnamefont {Livnat}}, \
  and\ \bibinfo {author} {\bibfnamefont {U.}~\bibnamefont {Zwick}},\ }\href
  {\doibase https://doi.org/10.1016/j.jalgor.2004.06.001} {\bibfield  {journal}
  {\bibinfo  {journal} {Journal of Algorithms}\ }\textbf {\bibinfo {volume}
  {53}},\ \bibinfo {pages} {169 } (\bibinfo {year} {2004})}\BibitemShut
  {NoStop}%
\bibitem [{\citenamefont {Farhi}\ and\ \citenamefont
  {Harrow}(2016)}]{farhi2016sampling}%
  \BibitemOpen
  \bibfield  {author} {\bibinfo {author} {\bibfnamefont {E.}~\bibnamefont
  {Farhi}}\ and\ \bibinfo {author} {\bibfnamefont {A.~W.}\ \bibnamefont
  {Harrow}},\ }\href@noop {} {\  (\bibinfo {year} {2016})},\ \Eprint
  {http://arxiv.org/abs/1602.07674} {arXiv:1602.07674 [quant-ph]} \BibitemShut
  {NoStop}%
\bibitem [{\citenamefont {Watts}\ and\ \citenamefont
  {Strogatz}(1998)}]{Watts1998}%
  \BibitemOpen
  \bibfield  {author} {\bibinfo {author} {\bibfnamefont {D.~J.}\ \bibnamefont
  {Watts}}\ and\ \bibinfo {author} {\bibfnamefont {S.~H.}\ \bibnamefont
  {Strogatz}},\ }\href {\doibase 10.1038/30918} {\bibfield  {journal} {\bibinfo
   {journal} {Nature}\ }\textbf {\bibinfo {volume} {393}},\ \bibinfo {pages}
  {440} (\bibinfo {year} {1998})}\BibitemShut {NoStop}%
\bibitem [{\citenamefont {Streif}\ and\ \citenamefont
  {Leib}(2020)}]{streif2019}%
  \BibitemOpen
  \bibfield  {author} {\bibinfo {author} {\bibfnamefont {M.}~\bibnamefont
  {Streif}}\ and\ \bibinfo {author} {\bibfnamefont {M.}~\bibnamefont {Leib}},\
  }\href {\doibase 10.1088/2058-9565/ab8c2b} {\bibfield  {journal} {\bibinfo
  {journal} {Quantum Science and Technology}\ }\textbf {\bibinfo {volume}
  {5}},\ \bibinfo {pages} {034008} (\bibinfo {year} {2020})}\BibitemShut
  {NoStop}%
\end{thebibliography}%
